\documentclass[%
 reprint,
 amsmath,amssymb,
 pre,
floatfix,
]{revtex4-1}

\usepackage{graphicx}
\usepackage{caption}
\usepackage{subcaption}
\usepackage{dcolumn}
\usepackage{bm}
\usepackage{epstopdf}
\usepackage{color}

\begin{document}

\preprint{APS/123-QED}

\title{Transition from 2D HD to 2D MHD turbulence}

\author{Kannabiran Seshasayanan}
 \email{skannabiran@lps.ens.fr}
\author{Alexandros Alexakis}%
\affiliation{%
 Laboratoire de Physique Statistique, {\'E}cole Normale Sup{\'e}rieure, CNRS UMR 8550, Universit{\'e} Paris Diderot, Universit{\'e} Pierre et Marie Curie, 24 rue Lhomond, 75005 Paris, France }%

\date{\today}

\begin{abstract}
We investigate the critical transition from an inverse cascade of energy 
to a forward energy cascade in a two-dimensional magneto-hydrodynamic flow 
as the ratio of magnetic to mechanical forcing amplitude is varied.
%
%
It is found that the critical transition is the result of two competing processes. 
The first process is due to hydrodynamic interactions, cascades the energy to the large scales.
The second process couples small scale magnetic fields to large scale flows
transferring the energy back to the small scales via a non-local mechanism.
At marginality the two cascades are both present and cancel each other. 
The phase space diagram of the transition is sketched.

\begin{description}
\item[PACS numbers]
May be entered using the \verb+\pacs{#1}+ command.
\end{description}
\end{abstract}

\pacs{Valid PACS appear here}
\maketitle


\section{Introduction}

The cascade of ideal invariants across scales,  is a fundamental concept of turbulence theory. 
In three dimensional flows energy and helicity cascade forward to the small scales while in
two dimensions energy cascades inversely to the large scales and enstrophy cascades forward 
to the small scales. 
Many flows in nature however \cite{byrne2013height}
show both characteristics: the formation of large structures due 
to an inverse cascade of energy and small scale turbulence due to a forward energy cascade.
Coexistence of forward and inverse cascades has been observed in laboratory and in numerical experiments  
under different physical situations: 
confined turbulence in thin layers,
rotating and stratified turbulence,
turbulence in the presence of strong magnetic fields,
2D turbulence in the presence of magnetic forcing
\citep{Celani10, Marino13, sozza2015dimensional, deusebio2014dimensional, Moubarak12, Xia11_1, Kanna14, Alexakis11}. 
In these studies the amplitude of the forward and inverse cascade can be varied by changing the relevant control parameter
(rotation, stratification, geometric factor, magnetic field strength etc). Due to high computational costs of three dimensional
numerical simulations, it is difficult to determine precisely the way the cascade transitions from forward to
inverse. A notable exception is the case of transition 
from two-dimensional         hydrodynamic turbulence (2D HD) 
to   two-dimensional magneto-hydrodynamic turbulence (2D MHD)  \cite{Kanna14}. 
In this case the transition is not ``dimensional" and all the simulations can be carried out 
in two dimensions which is computationally less expensive.  
For weak magnetic forcing the flow behaves as 2D HD cascading energy inversely. 
As the magnetic forcing is increased the inverse cascade decreases and finally stops at a critical amplitude
of the control parameter. Close to this critical point the flux of the inverse cascade scales as a power law with the distance from criticality.
This property manifests itself in the limit of large box size while at moderate box-sizes the transition appears smooth. 

In this work we unravel the mechanisms involved in this transition by looking 
in detail the different processes involved in transferring ideal invariants across scales
both in spectral and real space. 
In the next section  (\ref{discrip}) we describe in detail our system and define the non-dimensional control parameters
and observables that we are using in our analysis.
Section (\ref{CB}) presents the behavior of global (space and time averaged) quantities close to the critical point.
Section (\ref{FS}) shows how the fluxes and spectra vary as the control parameter is varied.
Section (\ref{SP}) discusses the effect of the transition on spatial structures and their statistics.
Finally, in the last section we summarize this work, sketch the phase space diagram and draw our conclusions.    

\section{Description of the system \label{discrip}}

The $2D$ MHD system is governed by the Navier-Stokes equation for the velocity field ${\bf u}$ and the induction equation for the magnetic field ${\bf b}$. In terms of the vorticity $\omega =\hat{\bf e}_z \cdot \nabla \times {\bf u}$ and the vector potential $a$, ${\bf b}=\nabla \left( \times { a \hat{\bf e}_z} \right)$, the governing equations read: 
\begin{eqnarray}
\partial_t \omega  + {\bf u \cdot \nabla} \omega   =&  
                {\bf b \cdot \nabla} j  &+ \nu^+ \nabla^{2n} \omega  + \nu^-  \nabla^{-2m} \omega, + \phi_\omega \nonumber \\
\partial_t a  + {\bf u \cdot \nabla} a  =&                          &+ \eta^+\nabla^{2n} a  + \eta^- \nabla^{-2m} a  + \phi_a
\end{eqnarray}
where $j=\hat{\bf e}_z \cdot \nabla \times {\bf b}$ is the current. $\nu^{+}, \eta^{+}$ are small-scale dissipation coefficients while $\nu^{-}, \eta^{-}$ are large-scale dissipation coefficients. The parameter $n, m$ give the order of the laplacian used in the dissipation terms.
The physically motivated values are $n=1$ and $m=0$. However, in the present work we use hyperviscosity to increase the inertial range and fix these values at $n=m=2$. These equations are numerically solved on a doubly periodic $2\pi L \times 2\pi L$ domain using a pseudospectral method. A standard Runge-Kutta of fourth order scheme is used for time marching (see \cite{Gomez05} for further details on the code).

\subsection{Nondimensional parameters}

The main control parameter in the present study is the ratio of the forcing in the magnetic field to the forcing in the velocity field $\mu_f=\frac{\| \bf F_b \|}{\| \bf F_u \| }$. ${\bf F_u}$ is the mechanical forcing given by ${\bf F}_u = -\Delta^{-1} \; \nabla \times \left( \phi_\omega \hat{\bf e}_z \right)$ and  ${\bf F_b}$ is the magnetic forcing given by ${\bf F}_b = \nabla \times \left( \phi_a \hat{\bf e}_z \right)$,
with  
$\phi_\omega=2f_0k_f \cos(k_fx)\cos(k_fy)$ and $\phi_a=\mu_f f_0 k_f^{-1} \sin(k_fx)\sin(k_fy)$. 
$k_f \, L$ gives the wave-number at which the system is forced, non-dimensionalized by the box size. 
The ratio of the dissipation coefficients results in two
Prandtl numbers: one for the large scale $Pm^- = \eta^-/\nu^-$ and the one for small scales $Pm^+ = \eta^+/\nu^+$. Both of these Prandtl numbers are set to unity $Pm^+ = Pm^- = 1$. 
The strength of turbulence is measured by the Reynolds numbers at small and large scales $Re^{+} = \left( f_0^{1/2} k_f^{1/2-2n} \right) /|\nu^{+}|, Re^{-} = \left( f_0^{1/2} k_f^{1/2+2m} \right) /|\nu^{-}|$
defined here based on the forcing amplitude. 
$Re^+$ determines the extend of the forward energy cascade while 
$Re^-$ determines the extend of the inverse energy cascade. In all runs $k_fL$ was chosen sufficiently large, so that the cascade does not reach the box-size and no large scale condensate is formed \cite{xia2008turbulence, chan2012dynamics, chertkov2007dynamics}.

An alternative control parameter to $\mu_f$ is the ratio of the injection rates $\mu_{\epsilon}$ 
defined as ratio of injection energy in ${\bf u}$ to the injection energy in ${\bf b}$, $\mu_{\epsilon} \equiv I_b/I_u$ with 
$I_b \equiv  \left\langle {\bf F_b \cdot b} \right\rangle$ and $I_u \equiv \left\langle {\bf F_u \cdot u} \right\rangle$ where the angular brackets stand for spatial and time average. 
This parameter might be more fitting to compare with more theoretical models like shell/EDQNM models or for forcing functions 
for which the energy injection rates is fixed. 
However for the forcing chosen in this work $\mu_\epsilon$ is not a control parameter in our system but an observable.

\subsection{Observables} 

In the ideal $2D$ hydrodynamic flow there are two conserved quantities in the system the kinetic energy 
                  $E_U    = \frac{1}{2} \left\langle |{\bf u}|^2\right\rangle_{_V}$
and the enstrophy $\Omega = \frac{1}{2} \left\langle      w^2   \right\rangle_{_V}$ (where $\left\langle \cdot \right\rangle_{_V}$ stands for spatial average). 
$E_U$ cascades to larger scales and $\Omega$ cascades to smaller scales \cite{boffetta2012two}.
On the other hand in the ideal 2D MHD system the two conserved quantities are the total energy 
$E = \frac{1}{2}\left\langle |{\bf u}|^2 + |{\bf b}|^2 \right\rangle_{_V}$ that cascades to the small scales and the square vector potential 
$A = \frac{1}{2}\left\langle a^2                       \right\rangle_{_V}$ that cascades to the large scales \cite{fyfe1976high, pouquet1978two}.
When forcing and diffusion is included in the system in the long time limit the system reaches a steady state where the
injection rate of all ideally-conserved quantities is balanced by their dissipation rate in the small and large scales. 
In this case choosing $\mu_f = 0$ reduces the system in the long time limit to a 2D HD flow since any initial magnetic field will 
disappear due to the anti-dynamo theorem for two-dimensional flows \cite{Zeldovich:1957zl}.
Accordingly energy will be dissipated in the large scales while enstrophy will be dissipated in the small scales.
For  non-zero values of $\mu_f $ the magnetic field will be sustained. 
However,
if the magnetic forcing is sufficiently small $\mu_f\ll 1$ the magnetic field will be too weak to feed back on the flow. 
In this limit the vector potential acts like a passive scalar advected by the 2D HD flow. 
In addition to $E_U$ and $\Omega$ the advection term also conserves the square vector potential $A$ that cascades to the small scales.
As the control parameter $\mu_f$ becomes larger, the Lorentz force eventually acts back to the flow
and magnetic field  stops being passive. Nonlinearities no longer conserve 
$\Omega$ and $E_U$ since the Lorentz force can inject/absorb kinetic energy and enstrophy to/from the flow.
The nonlinearities conserve the total energy $E$ and the square vector potential $A$.
At sufficiently large $\mu_f$ the system transitions to 2D MHD state that cascades the
total energy $E$  forward to the small scales while $A$ cascades inversely to the large scales. 

The strength of an inverse or a forward cascade of a quantity at steady state can be measured by
the rate of dissipation in the large and small scales respectively.
The rate of energy dissipation at large and small scales denoted by $\epsilon_{_E}^-, \epsilon_{_E}^+$ respectively are defined in Fourier space as, 
\begin{eqnarray}                                                                                                                                                      
\epsilon_{_E}^+ \equiv  &          |\nu^{+}| \left\langle \sum_{\bf k\ne0} |{\bf k}|^{ 2n} \left( |\tilde{\bf u}_{\bf k}|^2 + |\tilde{\bf b}_{\bf k}|^2  \right) \right\rangle_{_T}\nonumber \\ 
\epsilon_{_E}^- \equiv  & \;\;\;\; |\nu^{-}| \left\langle \sum_{\bf k\ne0} |{\bf k}|^{-2m} \left( |\tilde{\bf u}_{\bf k}|^2 + |\tilde{\bf b}_{\bf k}|^2| \right) \right\rangle_{_T}.
\end{eqnarray}
where $\left\langle \cdot \right\rangle_{_T}$ stands for time average. The rate of dissipation of the square vector potential in small and large scales denoted by $\epsilon_{_A}^+, \epsilon_{_A}^-$ are defined as
\begin{eqnarray}
\epsilon_{_A}^+ \equiv & |\nu^{+}|  \left\langle  \sum_{\bf k\ne0} |{\bf k}|^{ 2n} |a_{\bf k}|^2   \right\rangle_{_T} \nonumber \\
\epsilon_{_A}^- \equiv & |\nu^{-}|  \left\langle  \sum_{\bf k\ne0} |{\bf k}|^{-2m} |a_{\bf k}|^2   \right\rangle_{_T}
\end{eqnarray}
and similarly for the enstrophy dissipation rates $\epsilon_{_\Omega}^\pm$ we define
\begin{eqnarray}                                                                                                                                                      
\epsilon_{_\Omega}^+ \equiv  & |\nu^{+}|    \left\langle       \sum_{\bf k\ne0} |{\bf k}|^{ 2n+2}  |\tilde{\bf u}_{\bf k}|^2  \right\rangle_{_T}\nonumber \\ 
\epsilon_{_\Omega}^- \equiv  & \;\;\;\; |\nu^{-}| \left\langle \sum_{\bf k\ne0} |{\bf k}|^{-2m+2}  |\tilde{\bf u}_{\bf k}|^2 \right\rangle_{_T}.
\end{eqnarray}
Here ${\bf \tilde{u}_k, \tilde{b}_k}, \tilde{a}_{\bf k}$ denote the Fourier modes of ${\bf u, b}$ and $a$ respectively. 
The total dissipation of energy is given by $\epsilon_{_E} = \epsilon_{_E}^{-} + \epsilon_{_E}^{+}$, similarly for the square vector potential  $\epsilon_{_A} = \epsilon_{_A}^{-} + \epsilon_{_A}^{+}$ and vorticity $\epsilon_{_\Omega} = \epsilon_{_\Omega}^{-} + \epsilon_{_\Omega}^{+}$.

 The injection rate for the enstrophy is defined as $I_\omega \equiv \left\langle \omega \phi_\omega \right\rangle = I_{u} k_f^{2}$ and for the square vector potential as $I_{a} \equiv \left\langle  a \phi_a \right\rangle = I_b k_f^{-2}$.
Conservation laws then result to the relations $\epsilon_{_A} = I_a$ and $\epsilon_{_E} = I_u+I_b$. In the $\mu_f=0$ case we also have 
$I_{_\Omega}=\epsilon_{_{_\Omega}}$. 
%

In the limit of large Reynolds number $Re^+, Re^- \rightarrow \infty$ all injected conserved quantities are transported in scale space
from the forcing scale to the dissipation scales by a flux caused by the nonlinearity. 
Their dissipation rate is equal to the sum of their forward and inverse flux rates. 
The flux of kinetic energy $\Pi_{_U}$, enstrophy $\Pi_{_\Omega}$, total energy $\Pi_{_E}$ and 
square vector potential $\Pi_{_A}$ in the Fourier space are given by,
\begin{align} 
\Pi_{_U}(k) \equiv & \langle {\bf  u}_k^< {\bf   ( u\cdot\nabla u ) } \rangle \nonumber \\
\Pi_{_E}(k) \equiv & \langle {\bf  u}_k^< {\bf   ( u\cdot\nabla u  - b\cdot\nabla b)  +  b}_k^< {\bf ( u\cdot\nabla b - b\cdot\nabla u)}\rangle \nonumber \\
\Pi_{_A}(k) \equiv & \langle a_k^< ( {\bf u} \cdot\nabla a ) \rangle \nonumber \\
\Pi_{_\Omega}(k) \equiv & \langle {\omega}_k^< ( {\bf u} \cdot \nabla \omega ) \rangle 
\end{align}
where the notation $g^<_k$ represents the Fourier filtered field $g$ so that only the modes satisfying the condition $|{\bf k}| \le k$ is being kept, see \cite{Fris95}. 

Finally the distribution of energy among scales will be quantified through the kinetic $E_u$ and magnetic $E_b$ energy spectra:
\begin{align}
E_u(k)&=&\sum_{k\le |{\bf q}|<k+1} |{\bf u}_{\bf q}|^2 \\
E_b(k)&=&\sum_{k\le |{\bf q}|<k+1} |{\bf b}_{\bf q}|^2 .
\end{align}

\subsection{Numerical runs}

The aim of the present work is to reveal the mechanisms under which the cascade of the ideal invariants 
is changing as the control parameter $\mu_f$ is varied in the limit of large Reynolds numbers and box sizes. 
Due to computational limitations we can not have all three $Re^+, Re^-$ and $k_fL$ large for the same runs. 
As an alternative two sets of runs have been performed. In the first
for a moderate value of $Re^+$ the parameter
$\mu_f$ is varied for four different values of $k_fL$ and $Re^-$. The parameters for these sets of runs are shown in table \ref{Runs1}.
Part of the data from these runs were presented in \cite{Kanna14}. Here we present a more thorough analysis 
focusing on the mechanism that takes place during the observed transition.
For completeness some of the findings in \cite{Kanna14} are also shortly presented and discussed here.
In the second set of runs $\mu_f$ was varied for a fixed value of $k_fL=8$ and $Re^-$ and five different values of $Re^+$
to check the large $Re^+$ behavior. 
The parameters for these sets of runs are found in table
\ref{Runs2}. By comparing the different values of $Re^\pm$ tested we can verify if the examined runs have reached
a large $Re^\pm$ asymptotic behavior.
Finally, a third set of runs was performed with $k_fL=4$, $Re^+=1.4\,10^7$ and $Re^-=1.8\,10^3$, with focus on the 
development of energy spectra in the small scales. 

\begin{table}[!htb]
  \caption{Numerical parameters of the first set of DNS keeping $Re^+=1400$ fixed. $T$
expresses typical duration of the runs in units of $1/\sqrt{f_0 k_f}$.
}
   \begin{ruledtabular}
  \centering
    \begin{tabular}{c|ccccc} 
		Case     &  A1    & A2        &   A3      &    A4     \\ \hline
    		 $k_f$   &  8     &    16     &   32      &    64    \\ 
         $N$     & 512    &    1024    &  2048     &   4096          \\  
         $Re^-$  & $2.3\,10^4$  &  $7.4\,10^5$ &  $2.3\,10^6$ & $7.6\,10^7$  \\
         $Re^+$  & $1400$  &  $1400$ &  $1400$ & $1400$  \\
         $T$     & 2000   &   600     &   342   &  300       \\
    \end{tabular}
  \end{ruledtabular}
  \label{Runs1}
\end{table}

\begin{table}[!htb]
  \caption{Numerical parameters of the second set of DNS keeping $Re^-=23000, k_f \, L =8$ fixed. $T$ expresses typical duration of the runs in units of $1/\sqrt{f_0 k_f}$.  
  \label{Runs2}
}
   \begin{ruledtabular}
  \centering
    \begin{tabular}{c|ccccc} 
		Case     &  B1    & B2        &   B3      &    B4   &    B5       \\ \hline
         $k_f$   &  8     &    8     &   8      &    8      &    8        \\
         $N$     & 256    &    512    &  1024     &   2048  &  2048        \\  
         $Re^-$  & $23000$  &  $23000$ &  $23000$ & $23000$ & $23000$      \\
         $Re^+$  & $1.4\,10^2$  &  $1.4\,10^3$ &  $1.4\,10^4$ & $1.4\,10^5$ & $1.4\,10^6$    \\
         $T$     & 5000   &    2000    &   600     &   342   &  300       \\
    \end{tabular}
  \end{ruledtabular}
\end{table}

\section{Critical behavior \label{CB}}  

\subsection{Dimensional relations \label{dimrel}}

The relations between the injection rates, the forcing amplitudes and the amplitude of the
fluctuations can be constrained by simple dimensional and scaling arguments,
that we are using through out this paper.  
For large values of $Re^+$ the injection rates are expected to 
become independent of the dissipation coefficients in the small scales, and to 
depend only on the amplitude of the fluctuations at the forcing scale.
We obtain then
\begin{equation}
I_u \propto C_{_F} f_0 u_f  \quad  I_b \propto C_{_F} \mu_f f_0 b_f. \label{dim1} \end{equation} 
Here by $u_f$ and $b_f$ we denote the amplitude of the velocity and magnetic fluctuations 
respectably at the forcing scale.
In the presence of an inverse cascade the coefficient $C_{_F}$ can depend on the amplitude of the large scale fluctuations
that sweep the small scales and cause a fast decorrelation of the forcing and 
the velocity and magnetic fluctuations that reduces the injection rate \cite{tsang2009forced}.
Thus in general $C_{_F}$ is a function of $Re^-$ but independent of $Re^+$.
The injected energy, enstrophy and square vector potential is transferred to large or small scales 
by the nonlinearities. Assuming locality of interactions the fluxes and dissipation rates 
at small and large scales can be estimated to be 
\begin{equation}
\begin{array}{ccl}  
      \epsilon_{_E}     ^\pm &= \Pi_{_E}     ^\pm &\propto   u_\ell^3/\ell \\ 
      \epsilon_{_A}     ^\pm &= \Pi_{_A}     ^\pm &\propto   u_\ell b_\ell^2 \ell \\
      \epsilon_{_\Omega}^\pm &= \Pi_{_\Omega}^\pm &\propto   u_\ell^3/\ell^3
\end{array}
\label{dim2}
\end{equation}
where the relation for $\Omega$ only holds in the $\mu_f\ll1$ limit, and the 
relation $u_\ell^3/\ell$ should be replaced by $u_\ell(u_\ell^2+c b_\ell^2)/\ell$ when magnetic
effects become important. 
The proportionality coefficient in these relations is in general a non-trivial function of $\mu_f$
and they can become zero in the case of critical transitions.
This dependence on $\mu_f$ as well as the validity of the locality assumption can not be 
concluded from dimensional arguments and needs to be extracted from numerical simulations
and this is one of the main objectives of this work.
  
The dependence on $\mu_f$ however can be guessed in some limits.
For example in the case that $ \mu_f \ll1 $ the velocity field is not affected 
by the magnetic field and thus the usual relations of 2D HD turbulence hold
with $u_\ell \propto \Pi_{_\Omega}^{1/3} \ell$ at small scales and $I_u$ being independent of $\mu_f$. In the same limit using the two relations \ref{dim1} and \ref{dim2} at the forcing scale 
we obtain,
\begin{eqnarray}
\frac{b_f}{u_f} \propto \mu_f, \quad \mathrm{and} \quad \frac{I_b}{I_u} \propto \frac{\Pi_{_A} k_f^4}{\Pi_{_{\Omega}}} \propto \mu_f^2. \label{eqn:imp}
\end{eqnarray} 
The balance of the fluxes with the small scale dissipation rates also leads to 
the prediction of the dissipation length scales $\ell_\nu$ that are given by
$\ell_\nu \propto \ell_f\, (Re^+)^{3/(2-6n)}$ in the case the forward total energy cascade is dominant,
$\ell_\nu \propto \ell_f\, (Re^+)^{-1/2n}$ in the case the forward square vector potential cascade
or the enstrophy cascade is dominant. Implications of these relations have been tested in the sections that follow.

\subsection{The limit of $Re^-,k_fL \rightarrow \infty$}

\begin{figure}[!htb]
\includegraphics[scale=0.15]{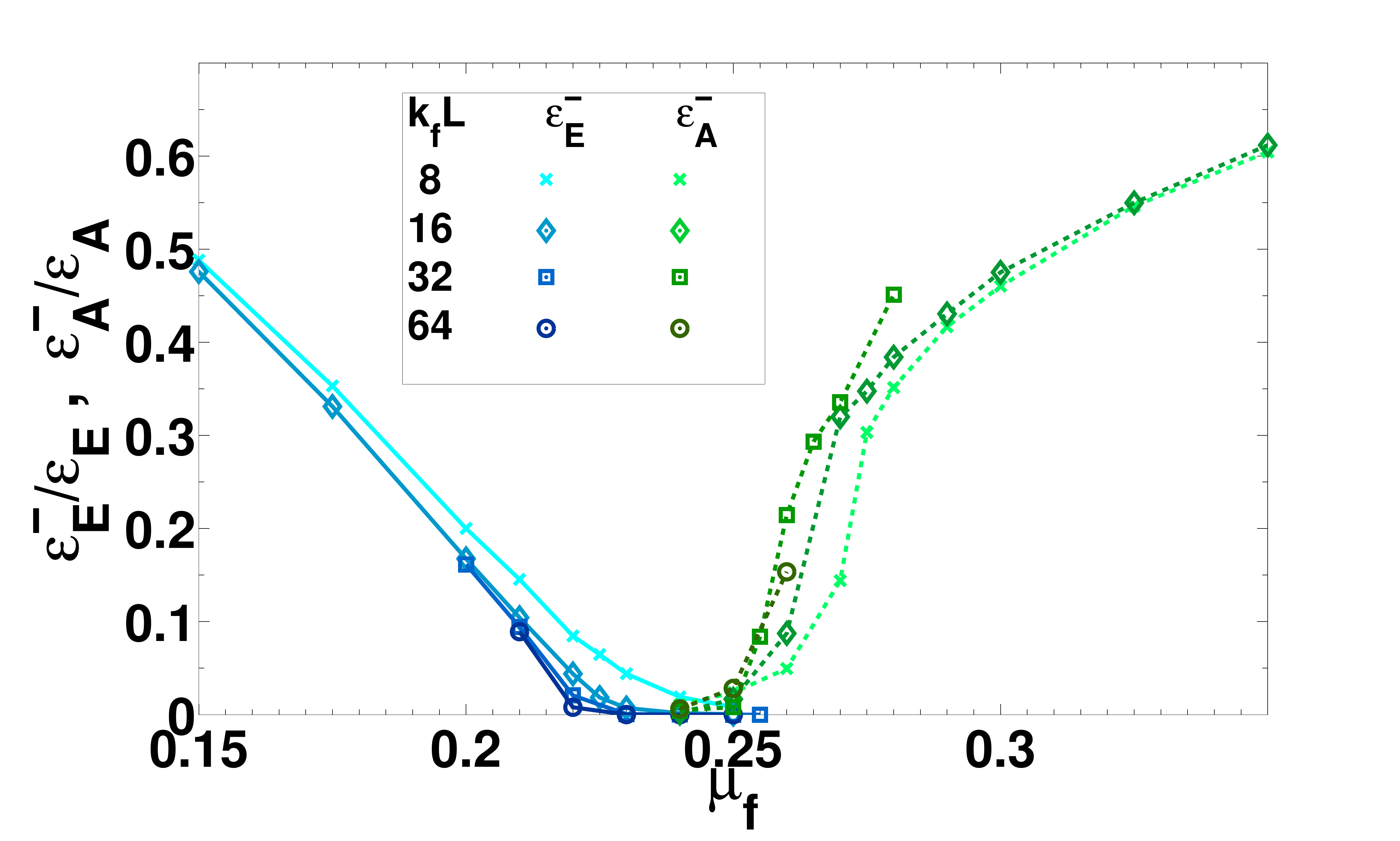}
\caption{\label{fig:transmuf} The normalized dissipation rates $\epsilon_{_E}^-/\epsilon_{_E}, \epsilon_{_A}^-/\epsilon_{_A}$ are shown as functions of $\mu_f$ for different values of $Re^{-}$ as mentioned in table \ref{Runs1}. Blue, continuous line corresponds to $\epsilon_{_E}^-/\epsilon_{_E}$ and green, dashed line to $\epsilon_{_A}^-/\epsilon_{_A}$ with darker shades denote larger $Re^-, k_f \, L$ as mentioned in the legend.}
\end{figure}

We begin by investigating the large $Re^-$ and large box size limit.
Figure \ref{fig:transmuf} shows the normalized dissipation rates $\epsilon_{_E}^{-}/\epsilon_{_E}$, and $\epsilon_{_A}^{-}/\epsilon_{_A}$ 
as functions of $\mu_f$ for the four different values of $k_f \, L, Re^-$ from the set of runs described in table \ref{Runs1}. 
Different symbols denote different values of $k_f\, L = \left( 8, 16, 32, 64 \right)$ as mentioned in the legend. 
As the domain size and $Re^{-}$ is increased the large scale energy dissipation $\epsilon_{_E}^{-}/\epsilon_{_E}$ goes to zero  
at a critical value of $\mu_{f}$ implying that in the large box limit the transition is  critical:
there is a value of the control parameter $\mu_f$ for which the amplitude of the inverse cascade becomes exactly zero.
Similarly the inverse cascade of the square vector potential goes to zero $\epsilon_{_A}^{-}/\epsilon_{_A} \rightarrow 0$ 
in the large box limit at a different critical value of $\mu_{f}$. The two critical values of $\mu_f$ that are observed 
are denoted as $\mu_{_Ec}\simeq 0.22$  for which the inverse cascade of energy stops,
and the other for which the inverse cascade of square vector potential starts denoted as $\mu_{_Ac}\simeq 0.26$ . 
Between these two critical values no inverse cascade exists and both conserved quantities cascade only forward \cite{Kanna14}.

\begin{figure}[!htb]
\includegraphics[scale=0.15]{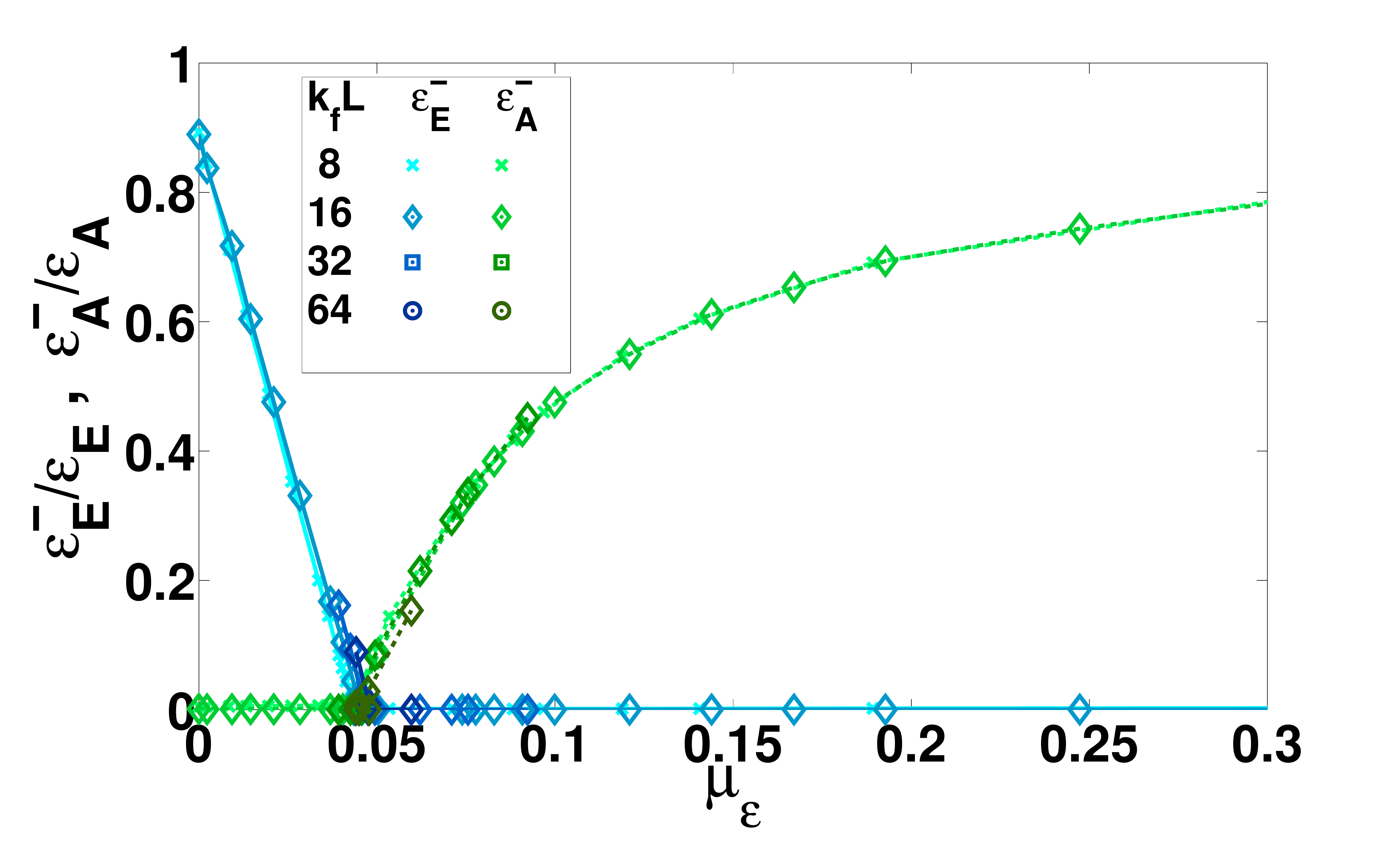}
\caption{\label{fig:transmue}  $\epsilon_{_E}^-/\epsilon_{_E}, \epsilon_{_A}^-/\epsilon_{_A}$ are shown as functions of $\mu_{\epsilon}$ for different values of $Re^{-}$ as mentioned in table \ref{Runs1}. Blue, continuous line corresponds to $\epsilon_{_E}^-/\epsilon_{_E}$ and green, dashed line to $\epsilon_{_A}^-/\epsilon_{_A}$ with darker shades denote larger $Re^-, k_f \, L$ as mentioned in the legend. }
\end{figure}

\begin{figure}[!htb]
\includegraphics[scale=0.15]{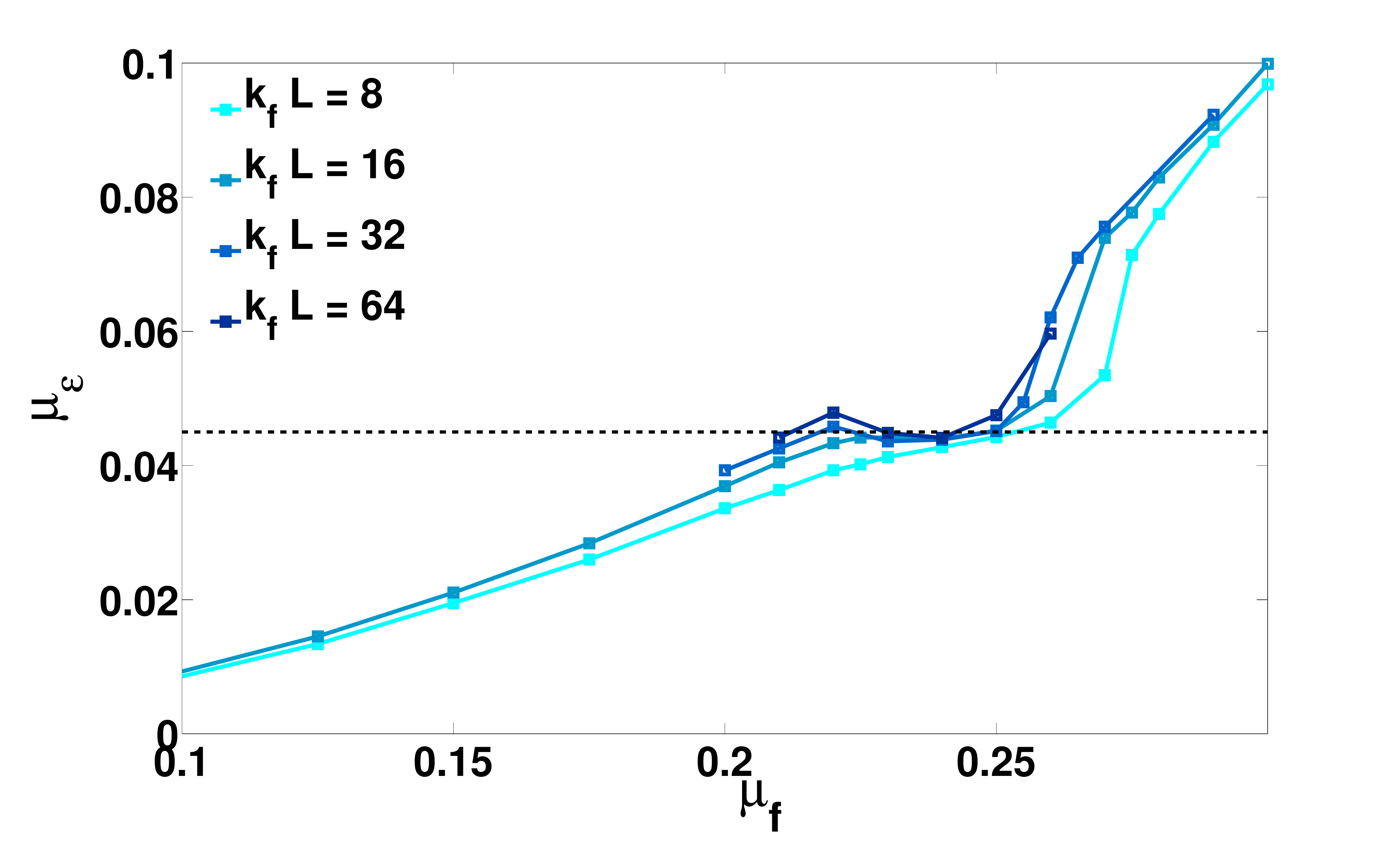}
\caption{\label{fig:mufmue} $\mu_{\epsilon}$ is shown as a function of $\mu_f$ for different values of $Re^{-}, k_f \, L$ for runs in table \ref{Runs1}. Darker shades denote larger $Re^{-}, k_f \, L$ as mentioned in the legend.}
\end{figure}

As discussed in the previous section 
the alternative control parameter $\mu_{\epsilon}$ can be used that is closer to theoretical models.
The normalized dissipation rates as a function of $\mu_\epsilon$ are shown in figure \ref{fig:transmue}. 
The shape of the transition curve appears to change when  the same data are presented as a function of $\mu_\epsilon$.
First the two critical points appear to merge to one, denoted now as $\mu_{\epsilon c}\simeq0.045$ . 
Second the two exponents appear that can be fitted 
close to the critical value are different from the $\mu_f$ case and are both closer to one.
These points imply a non-trival relation between  $\mu_\epsilon$ and $\mu_f$.
This relation is plotted in figure \ref{fig:mufmue} where the observable $\mu_{\epsilon}$ is shown as a function of $\mu_f$ 
for different values of $Re^-$. 
Close to the critical point $\mu_{\epsilon}$ becomes a non-monotonic function of $\mu_f$. 
This non-monotonic behavior becomes more pronounced as we increase the domain size and 
is localized around the critical points in the system.
As a result for the critical value of $\mu_{\epsilon}$ we can find three values of $\mu_f$
that satisfy $\mu_\epsilon(\mu_f)=\mu_{\epsilon c}$. Two of these values are the critical
values $\mu_{_Ec}$ and $\mu_{_Ac}$ found in figure \ref{fig:transmuf}, and the third to a value in between.

The parameter $\mu_{\epsilon}$ is thus a parameter worth considering for future studies for which the energy injection
rate is fixed rather the forcing amplitude. However at the present study we will
stay with $\mu_f$ description, since the uncertainty that comes from the measurement of $\mu_\epsilon$ 
makes it difficult to interpret the results close to the critical point.

\subsection{The limit of $Re^+ \rightarrow \infty$}\label{limit2}
\begin{figure}[!htb]
\begin{center}
\begin{subfigure}[b]{0.4\textwidth}
\includegraphics[scale=0.18]{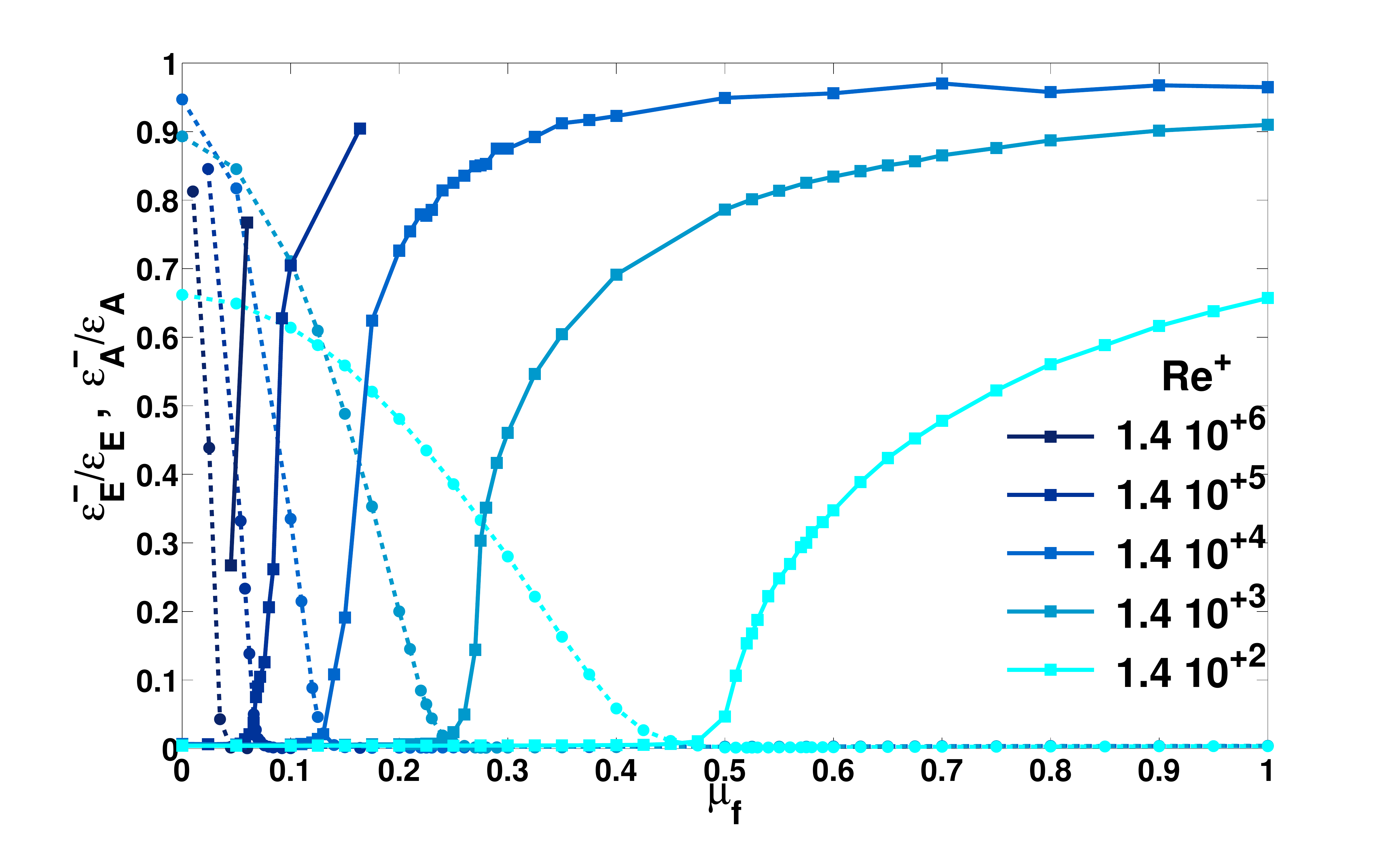}
\caption{}
\label{fig:firsttransR+}
\end{subfigure}
\begin{subfigure}[b]{0.4\textwidth}
\includegraphics[scale=0.18]{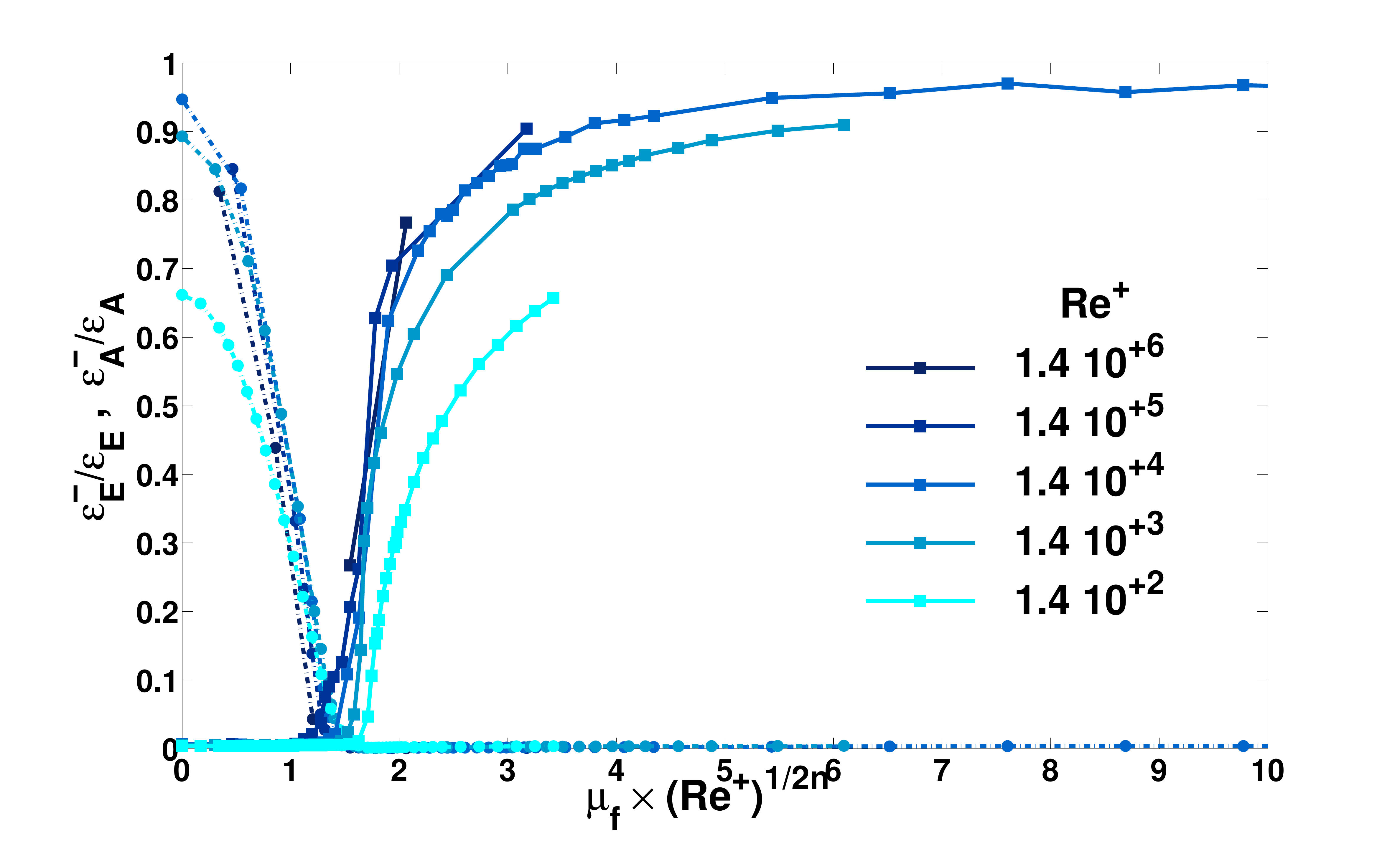}
\caption{}
\label{fig:secondtransR+}
\end{subfigure}
\end{center}
\caption{\label{fig:transR+} The top plot shows $\epsilon_{_E}^-/\epsilon_{_E}$ and  $\epsilon_{_A}^-/\epsilon_{_A}$ as functions of $\mu_f$ 
for different values of $Re^+$ mentioned in the legend for runs in table \ref{Runs2}. Dotted line represents $\epsilon_{_E}^-/\epsilon_{_E}$ 
while the continuous represents $\epsilon_{_A}^-/\epsilon_{_A}$. Darker shades denote larger values of $Re^+$ as mentioned in the legend. The plot below shows the same data with a rescaled $x$ axis $\mu_f \, \left( Re^+ \right)^{1/2n}$}
\end{figure}

The other limit considered is the limit of large $Re^+$ for fixed $k_f \, L=8$ and $Re^{-}=23000$
(see table \ref{Runs2}). A series of runs for five different values of $Re^{+}$ are made. 
For these set of runs the box size and $Re^-$ are not large enough for a clear manifestation of criticality
(that appears only in the large $k_f \, L$ and $Re^-$ limit).
For this reason we cannot distinguish between the two critical points $\mu_{_Ec}$ and $\mu_{_Ac}$ that appear as one
and we will simply refer to it as $\mu_c$.
Figure \ref{fig:firsttransR+} 
shows the normalized dissipation rates at large scales as a function of $\mu_f$ for different $Re^{+}$. 
The critical point moves to smaller values of the control parameter as we increase $Re^{+}$. 
However, rescaling the control parameter $\mu_f$ by multiplying with $\left( Re^{+} \right)^{1/2n}$
makes the curves collapse on top of each other as shown in Figure \ref{fig:secondtransR+}. 
This is not so surprising since the magnetic field 
is being stretched and amplified at the smallest scales and its amplitude depends on the viscous cut-off length-scale. 
The rescaling factor can be derived 
assuming that the transition takes place when the magnetic field and the velocity field are of the same strength. 
From standard phenomenological arguments for constant  flux (see section \ref{dimrel}) we have,
\begin{eqnarray}        
 a_{\ell}^2 \propto \Pi_{_A} \frac{\ell}{u_{\ell}} \propto \Pi_{_A} \frac{1}{\Pi_{_\Omega}^{1/3}} \label{eqn:scales_1}
\end{eqnarray}
The magnetic field at the viscous/ohmic scale $\ell_\nu$ is then given by, 
$b_{\nu} \propto a_{\nu} \ell_{\nu}^{-1} \propto u_f \left( \frac{\Pi_{_A}}{\Pi_{_\Omega}} \right)^{1/2} \frac{1}{\ell_{\nu} \ell_f}$.
Finally the small scale dissipation length scale $\ell_\nu$ can be estimated by balancing the flux of $A$ 
with its ohmic dissipation rate that leads
to $\ell_{\nu} \propto \ell_f \left( Re^{+} \right)^{-1/2n}$.
Assuming that the transition takes place when $b_{\nu} \sim u_f$, and using equation \ref{eqn:imp} we obtain
\begin{eqnarray}
\mu_c \propto \frac{\ell_{\nu}}{\ell_f} \propto \left( Re^+ \right)^{-1/{2n}}
\end{eqnarray}
This leads to the observed scaling for the critical point $\mu_c$ as a function of the forward Reynolds number.  
It is important here to note that in this scaling argument the transition occurs when there is an equality between the magnetic field at small scales 
and the velocity field at large scales. The transition is thus controlled by a nonlocal interaction between 
the small magnetic scales and the large velocity scales. 

\subsection{Injection rates and fluctuations}  

\begin{figure}[!htb]
\includegraphics[scale=0.15]{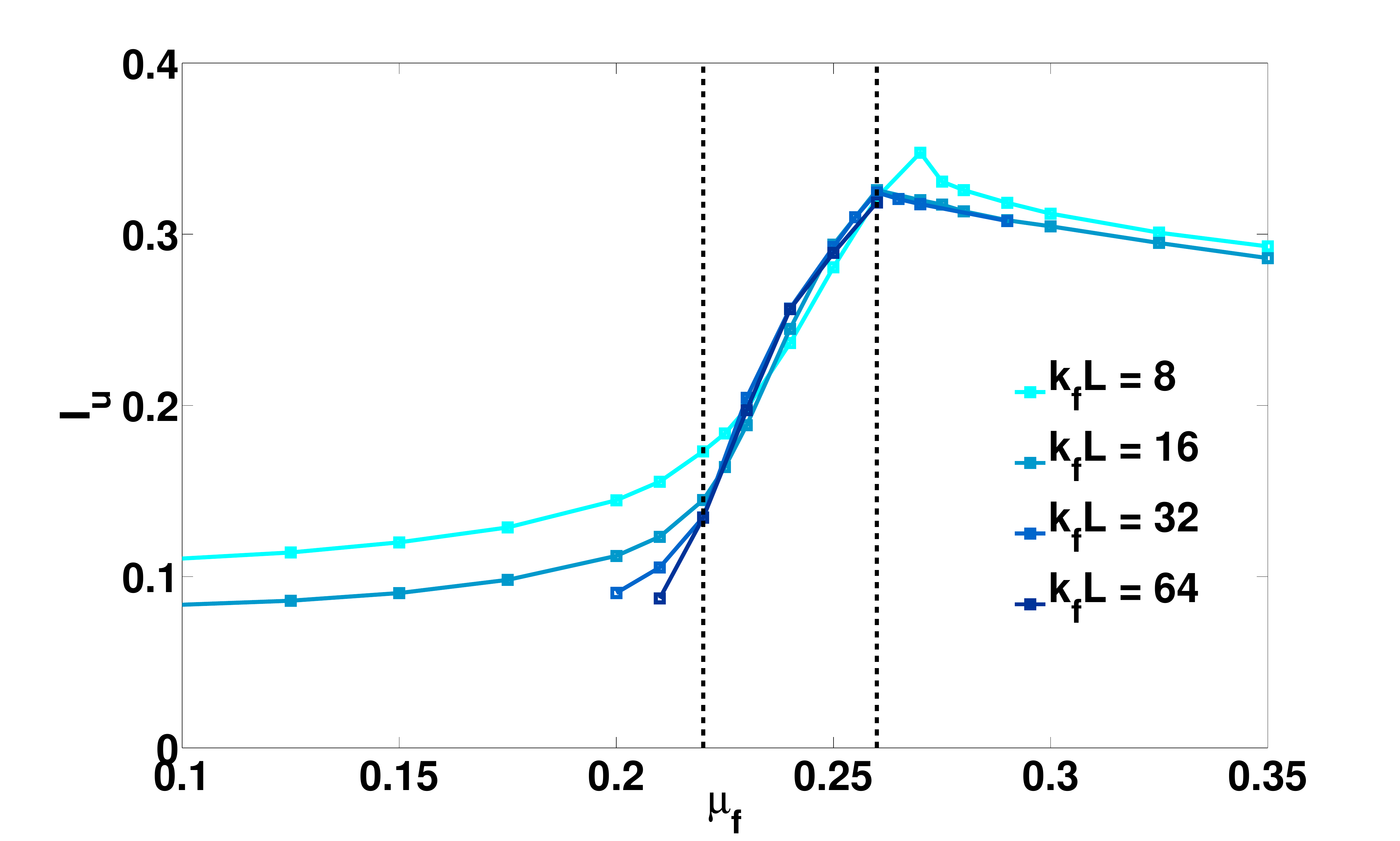}
\includegraphics[scale=0.15]{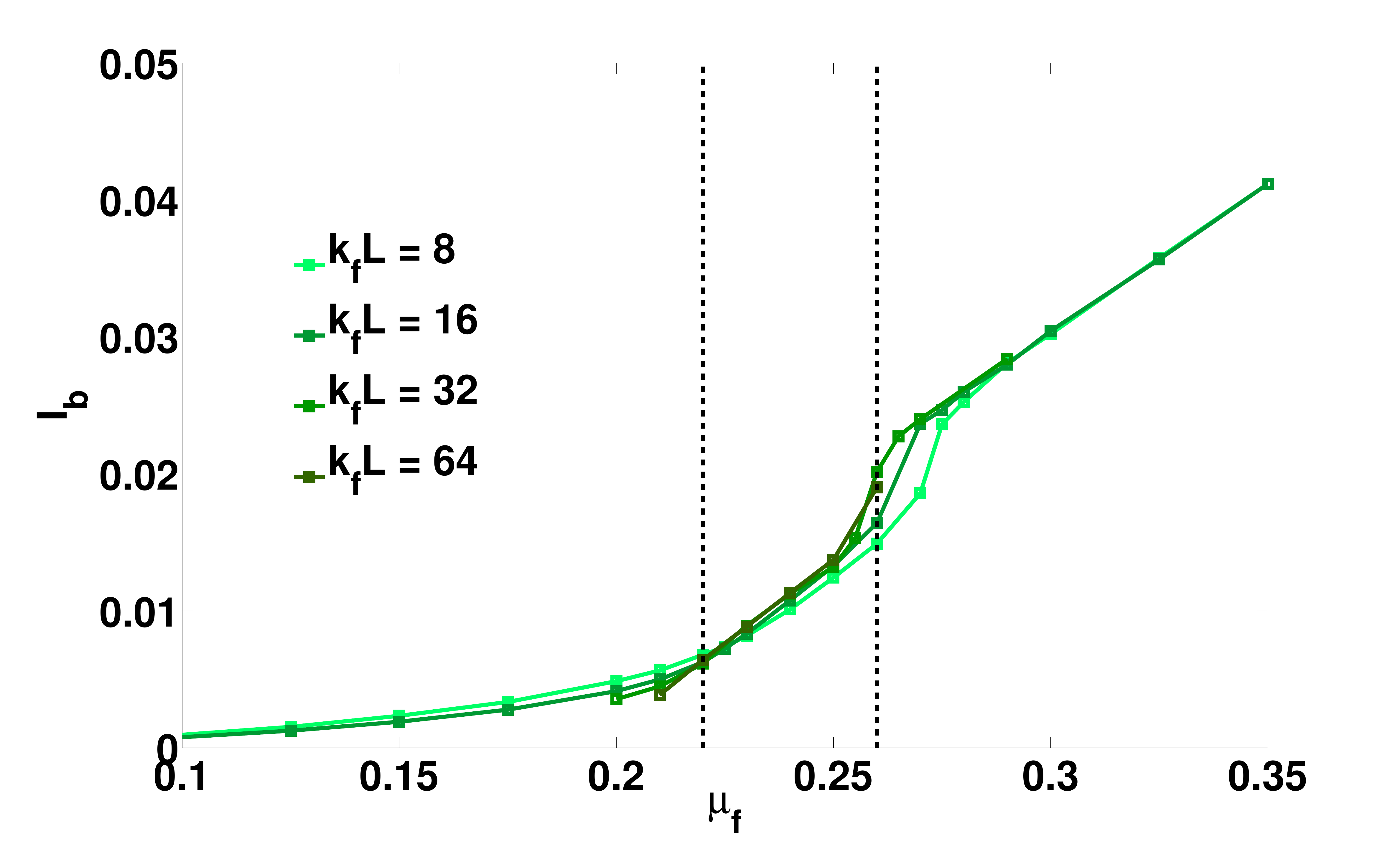}
\caption{\label{fig:forc} Figure shows on top $I_u$ and on bottom $I_b$, the injection energies in the kinetic and the magnetic field 
as functions of $\mu_f$. Darker shades denote larger values of $k_f \, L, Re^-$ as mentioned in the legend for runs in table \ref{Runs1}. The black lines denote values $\mu_f = 0.22, 0.26$ denoting proximity to the critical points $\mu_c$.}
\end{figure}

The relation between the injection rates and the amplitude of fluctuations is the most basic relation in steady state turbulence. 
The behavior of these quantities as the control parameter approaches close to the critical point is of primary interest
as they allow to estimate the rate energy cascades to the dissipation scales. 
Shown in figure \ref{fig:forc} are the 
injection of kinetic and magnetic energies $I_u=I_{\omega}k_f^{-2}$ and $I_b=I_ak_f^2$ as functions of $\mu_f$ for runs in table \ref{Runs1}.
We note that
$\mu_f$ was varied by keeping $F_u$ fixed and changing the value of $F_b$. 
Increasing values of $Re^-, k_f \, L$ are denoted with increasing shade of color of the lines.
Four values of $Re^-$ are shown.
The two vertical black lines at $\mu_f = 0.22, 0.26$  denote the proximity to the critical points. 

In the limit $\mu_f \ll 1$ the quantity $I_u$ reaches an asymptotic value since the magnetic field plays a passive role
and the flow is close to 2D HD. The value at 
the forcing scale is suppressed as $k_f \, L$ increases due to sweeping effect of the large scales from the 
inverse cascade \cite{tsang2009forced}. 
As expected the magnetic injection rate $I_b$ approaches zero in the $\mu_f\to0$ limit as $I_b\propto \mu_f^2$ see section \ref{dimrel}. 
As we increase $\mu_f$ both injection rates $I_u,I_b$ increase.
Close to the first critical point $\mu_{_Ec}$ both injection rates vary smoothly.
Between the two critical points the injection rates sharply increase and at the second 
critical point $\mu_{_Ac}$ both injection rates vary continuously but their first 
derivative with respect to $\mu_f$ appears to have a discontinuous change.
At large values of $\mu_f$ the magnetic energy injection rate further increases while the 
kinetic energy injection rate is suppressed.  Note however that for values of $\mu_f$ shown in figure \ref{fig:forc}, $I_b$ is always smaller than $I_u$.
It is also worth pointing out that if instead of
$\mu_f$ the data were plotted using $\mu_\epsilon$ both injection rates would appear
discontinuous at the critical point.   

\begin{figure}[!htb]
\includegraphics[scale=0.15]{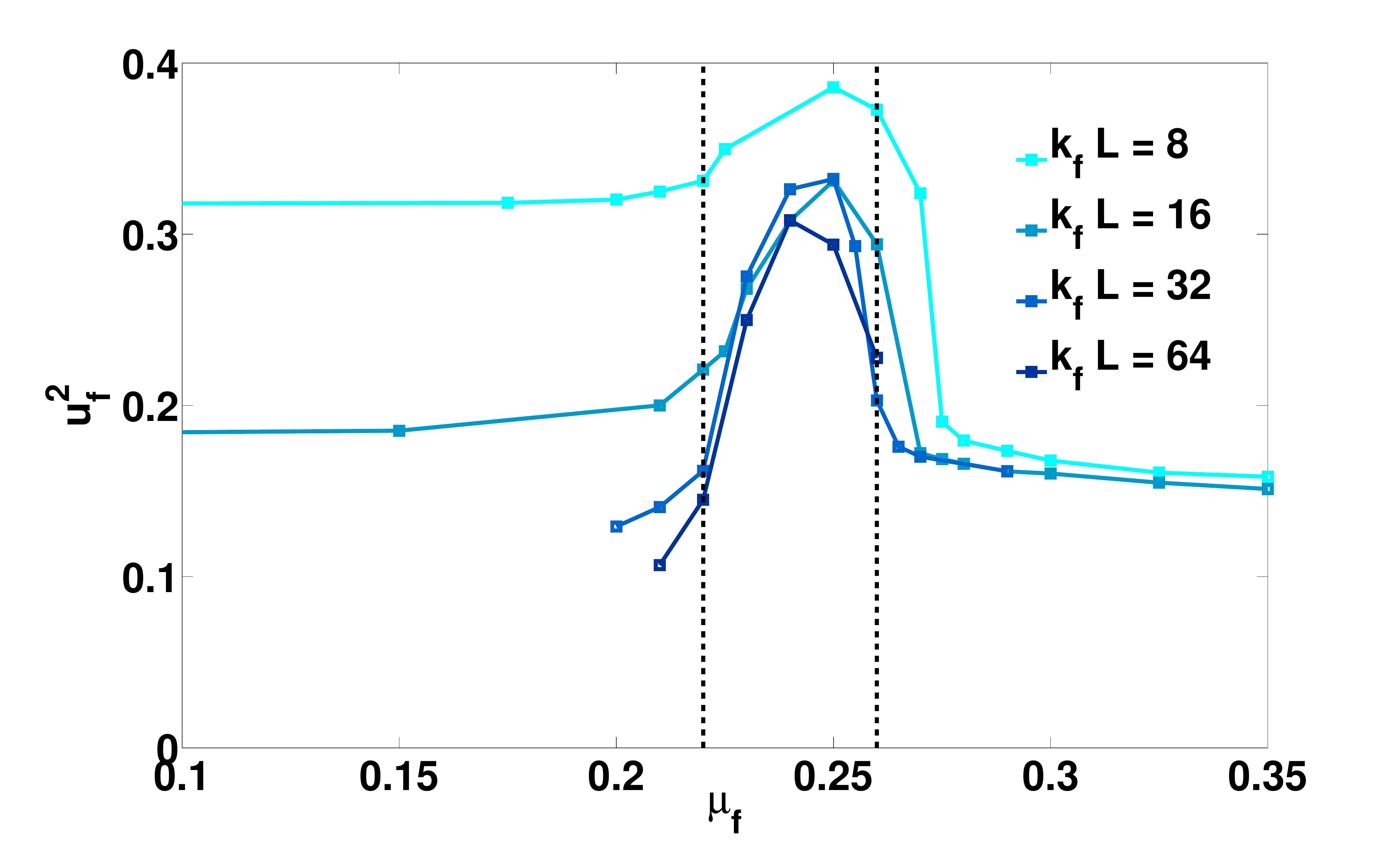}
\includegraphics[scale=0.15]{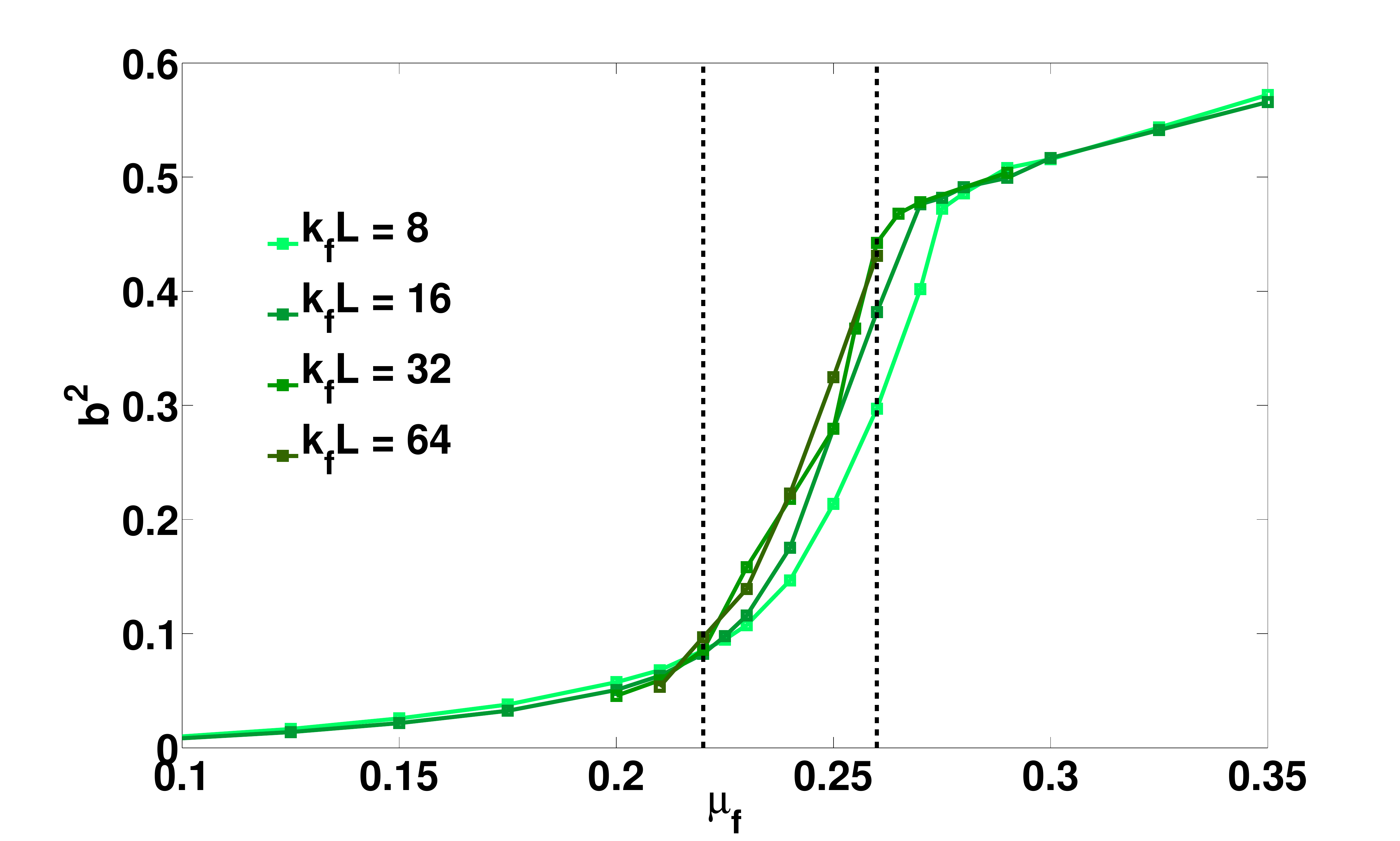}
\caption{\label{fig:forcvals} Figure shows on top $ u_f^2 $ and the bottom $ b^2 $,
representing the square of the velocity field at the forcing length scale $k_f$ and twice the magnetic energy as functions of $\mu_f$. Darker shades denote larger values of $k_f \, L, Re^-$ as mentioned in the legend for runs in table \ref{Runs1}. The black lines denote values $\mu_f = 0.22, 0.26$ denoting proximity to the critical points $\mu_c$.}
\end{figure}

The injected energy at the forcing scale is transported in the dissipation scales by nonlinear interactions.
As discussed before the transition occurs when $b^2$ becomes of the same order as $u_f^2$. Here $b^2$ is twice the total magnetic energy, $b^2 = \left\langle |{\bf b}|^2 \right\rangle$. Thus close to the transition the dominant role for the energy transport at the forcing scale is played by the velocity fluctuations 
at the same scale as the forcing and by the magnetic field fluctuations at the small viscous scales where the magnetic field is strong. 
The magnetic field at the forcing scale $b_f^2$ is weaker than $b^2$ by a factor $(Re^+)^{-1/n}$ as discussed in section \ref{limit2}. 
In figure \ref{fig:forcvals} we plot  $ u_f^2 $ 
the kinetic energy at the forcing scale 
(defined as the energy of the Fourier modes with $k=|{\bf k}|$ in the range $[k_f-1,k_f+1]$) 
and the magnetic energy $ b^2 $ that is dominant in the small scales.   
$ u_f^2 $
becomes maximum between the critical points where both the inverse cascade of energy and square vector potential vanishes, 
suggesting a pile up at the forcing scale $k_f$. The magnetic energy $ b^2 $ is much smaller than 
$u_f^2$ before the transition. 
$ b^2 $ 
increases smoothly and rapidly after the first critical point. 
Close to the second critical point there is a sharp change for both quantities.
$ b^2 $ shows a discontinuous first derivative
while the derivative of $ u_f^2 $ could also possibly diverge (from the right)
as $\mu_f$ is decreased to the $\mu_{_Ac}$ value.

\subsection{Balance relations}  

Having established the general behavior of the injection rates and the amplitude of fluctuations close to the critical points
we can attempt to give a phenomenological description of the cascade process.
In 3D turbulence and in the limit of small dissipation coefficients  one can argue by dimensional reasoning alone that the energy dissipation rate $\epsilon_{_E}$ will be proportional to the cubic power of the 
fluctuation amplitude and inversely proportional to the forcing length scale $\epsilon_{_E}=C_{_E} u_f^3/\ell_f$. 
The non-dimensional proportionality coefficient $C_{_E}$ can in principle depend 
on the forcing functional form. The same scaling holds for 2D turbulence but for the dissipation rate in the large scales
$\epsilon^-_{_E}$.

%
%

\begin{figure}[!htb]
\includegraphics[scale=0.15]{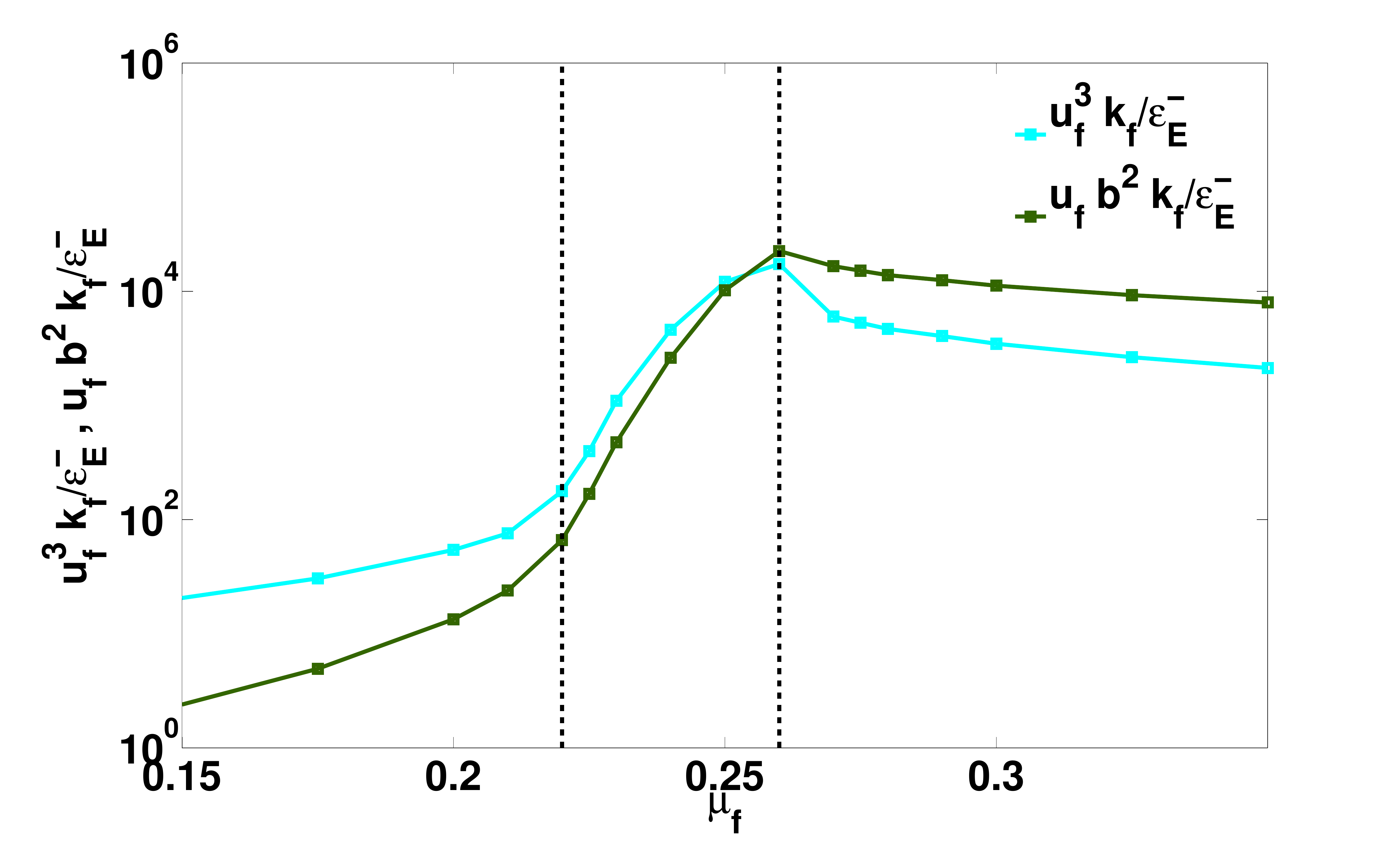}
\caption{\label{fig:Kolcomp} The quantities $u_f^3 k_f/\epsilon_{_E}^{-}$ and $u_f b^2 k_f/\epsilon_{_E}^{-}$ 
are shown as functions of $\mu_f$ for the case $k_f \, L = 16$, see table \ref{Runs1}.}
\end{figure}

The situation is different for the 2D MHD case, because there are two processes involved: 
vortex-shearing  that amplifies energy in the large scales \cite{chen2006physical},
and magnetic field line stretching that removes energy from the large scale vortices and amplifies energy in the small scales. 
Thus in principle we need to define two terms. 
The first one responsible for the inverse cascade is proportional to the cubic power of the fluctuations at the forcing scale $u_f^3 \, k_f$. 
The second one that couples small scale magnetic field to the forcing scale shear is proportional to $u_fb^2 \, k_f$ 
and suppresses the energy transport to the large scales. 
We thus expect for the inverse cascade:
\begin{equation}                                                                                                                                                      
\epsilon_{_E}^- = C_{_K}^- ( u_f^3k_f - \alpha \, u_f \, b^{2}k_f) 
\end{equation}
where $C_{_K}^-$ and $\alpha $ are non-dimensional constants that need to be determined from the data. 
As a proof of concept we plot in figure \ref{fig:Kolcomp} the quantities 
$u_f^3k_f/\epsilon_{_E}^{-}$ and $u_f b^2 k_f /\epsilon_{_E}^{-}$.  
For clarity only the case $k_f \, L = 16$ is shown. 
Indeed these two curves show that close to the critical points
the term responsible for the forward cascade $u_f b^2 k_f$ becomes of the same order as the term for the inverse cascade $u_f^3 k_f$.
For values $\mu_f \ll \mu_{_Ec}$ the term $u_f^3 k_f$ dominates while for $\mu_f \gg \mu_{_Ac}$ the term  $u_f b^2 k_f$  is bigger.
\begin{figure}[!htb]
\includegraphics[scale=0.15]{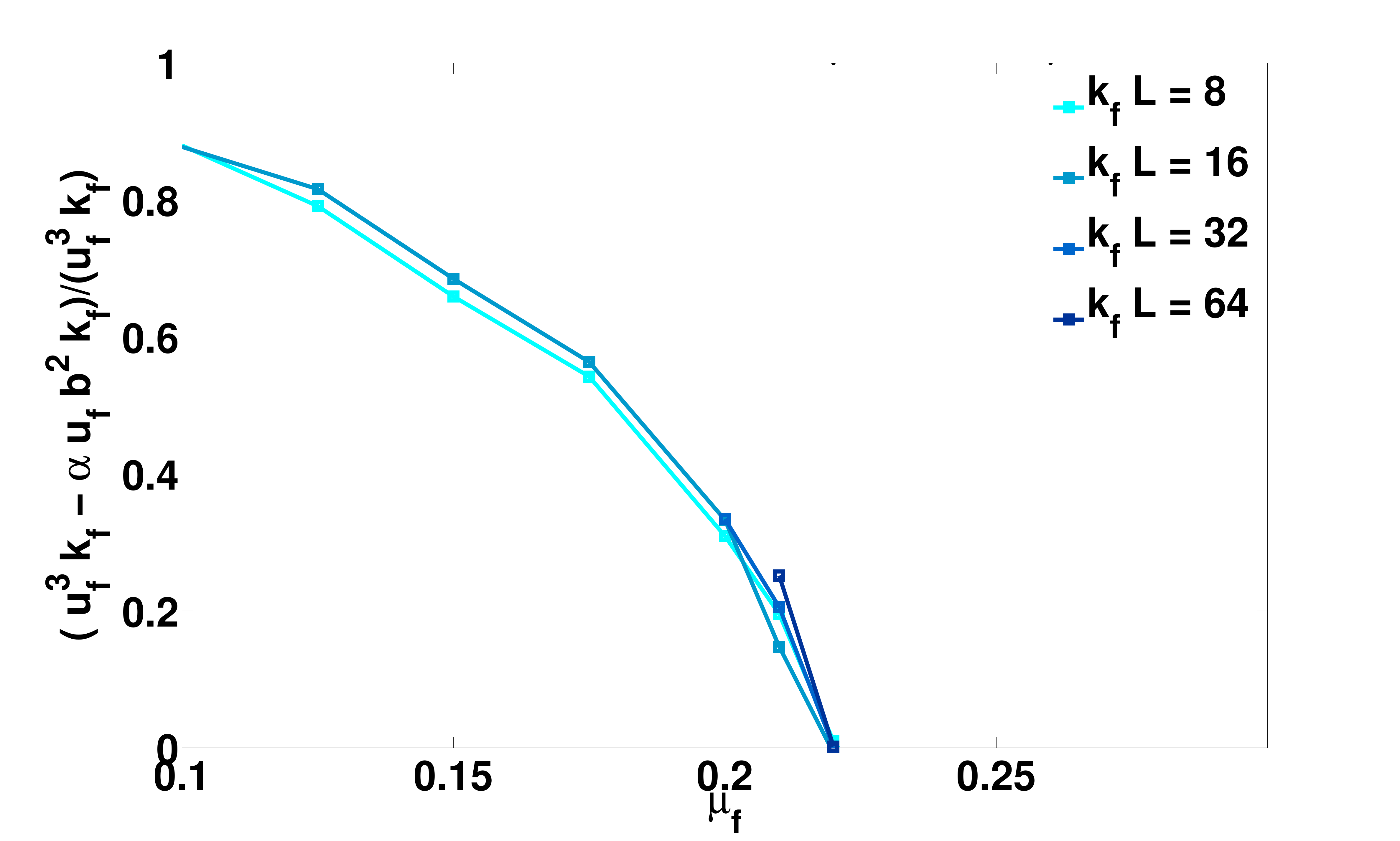}
\caption{\label{fig:uuufuub} The ratio $(u_f^3 k_f - \alpha u_f b^2k_f)/(u_f^3 k_f)$ is shown as a function of $\mu_f$. Darker shades denote larger values of $k_f \, L, Re^-$ as mentioned in the legend for runs in table \ref{Runs1}.}
\end{figure}
In figure \ref{fig:uuufuub} we plot the ratio $(u_f^3k_f - \alpha u_f b^2k_f)/(u_f^3 k_f)$ for the different values 
of $\mu_f$ and $Re^-$ of the runs presented in table \ref{Runs1}. The coefficient $\alpha$ has been chosen 
so that the numerator of this ratio becomes zero at the critical point. The data collapse to a single curve
indicating that the functional form of this curve is independent of the large scale parameters $Re^-$ and $k_fL$.
We note however that the power-law behavior close to the critical point 
in figure \ref{fig:uuufuub} differs from the one observed in figure \ref{fig:transmuf}. This indicates that the coefficient $C_{_K}$ also depends on $\mu_f$ or alternatively mean square values (such as $b^2, u_f^2$) cannot provide an accurate description near the transition and higher order statistics need to be taken into account. 

\begin{figure}[!htb]
\includegraphics[scale=0.15]{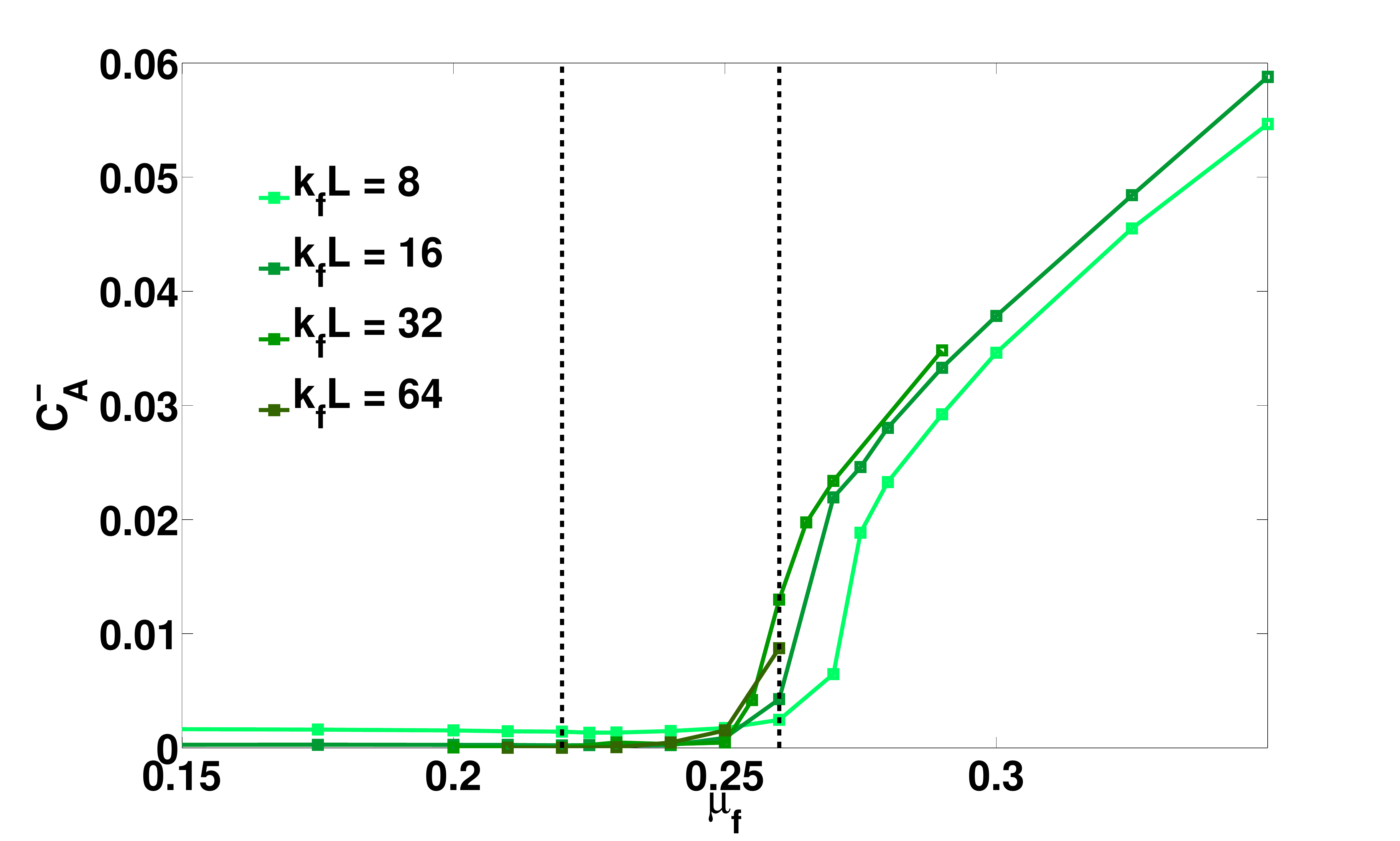}
\caption{\label{fig:Kolmogr} The Kolmogorov constant $C_{_A}^-$ is shown as a function of the control parameter $\mu_f$ for different values of $Re^-$, see table \ref{Runs1}. Vertical black lines are $\mu_f = 0.22, 0.26$ denoting proximity to the critical points $\mu_f$. Darker shades denote larger values of $k_f \, L, Re^-$ as mentioned in the legend.}
\end{figure}

The case for the inverse cascade of square vector potential is different because it is not 
as easy to identify the different processes involved. 
The simplest possible dimensionally acceptable form for the dependence of the inverse cascade
on the amplitude of the fluctuations is given by:  
\begin{equation}                                                                                                                                                      
\epsilon_{_A}^- = C_{_A}^- u_f a_f^{2} k_f =  C_{_A}^- u_f b_f^{2} k_f^{-1}
\end{equation}
where now $C_{_A}^-$ is a function of the control parameter $\mu_f$. 
Plotting the ratio $ \epsilon_{_A}^-/u_f b_f^{2} k_f^{-1} $ for all the data of the runs in table \ref{Runs1}
indeed collapses the curves to a single one, shown in figure \ref{fig:Kolmogr}. This indicates that the transition to the inverse cascade of the square vector potential
can be expressed by a ``Kolmogorov-coefficient"  that varies with $\mu_f$ and becomes zero at the critical point.
Note however that this simplification originates from our lack of knowledge of the precise mechanisms involved in the transition for square vector potential.
In principle even this transition could result from the competition of two processes (one forward e.g. by eddy stretching and one inverse
e.g. by a negative turbulent diffusivity) that equate at a critical value of $\mu_f$. However, at present we cannot 
identify these processes from the attained data, and leave such considerations for future work.

%

\section{Ideal invariants across scales \label{FS}}    

\subsection{Flux of conserved quantities}
\begin{figure}[!htb]
\includegraphics[scale=0.15]{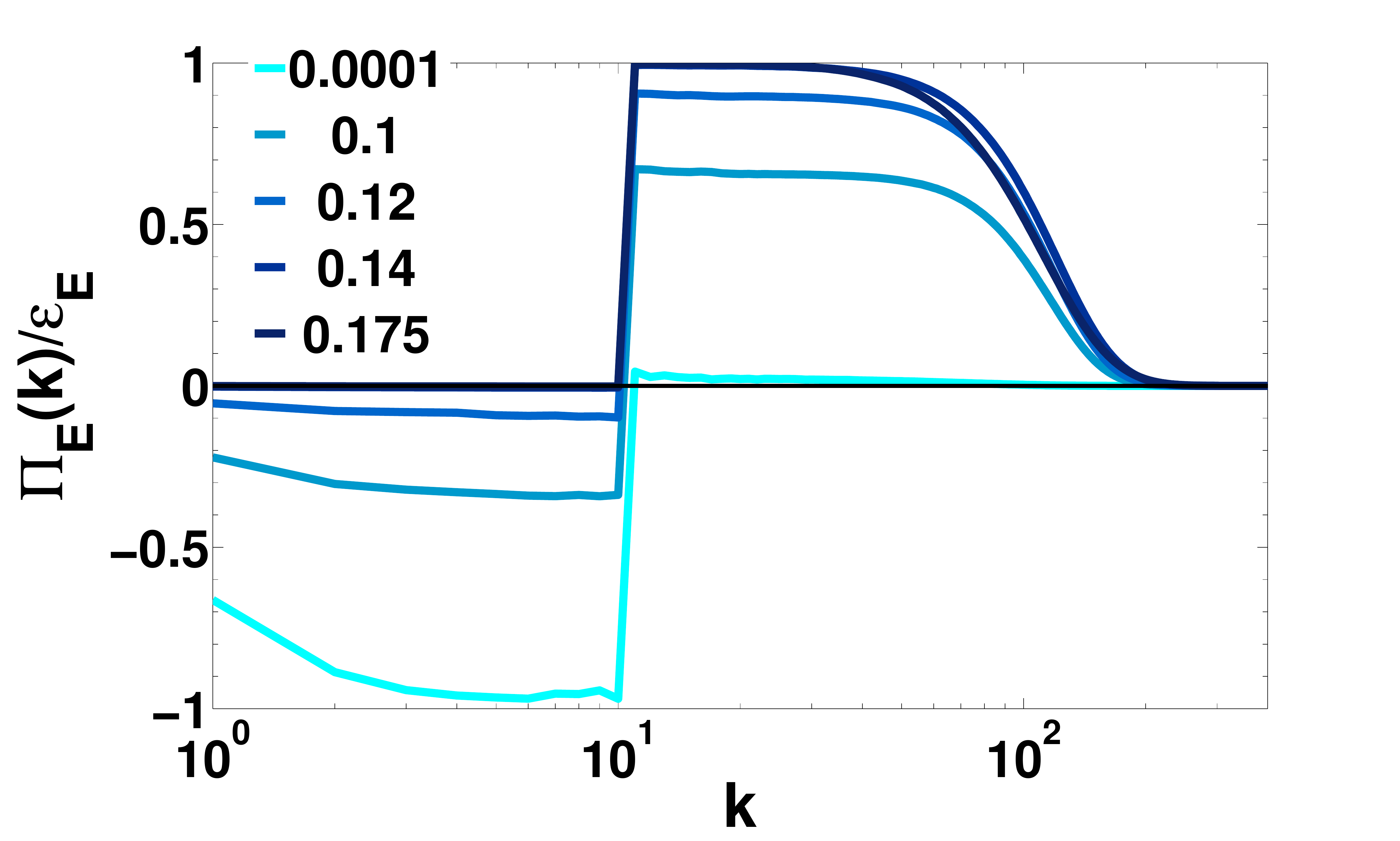}
\includegraphics[scale=0.15]{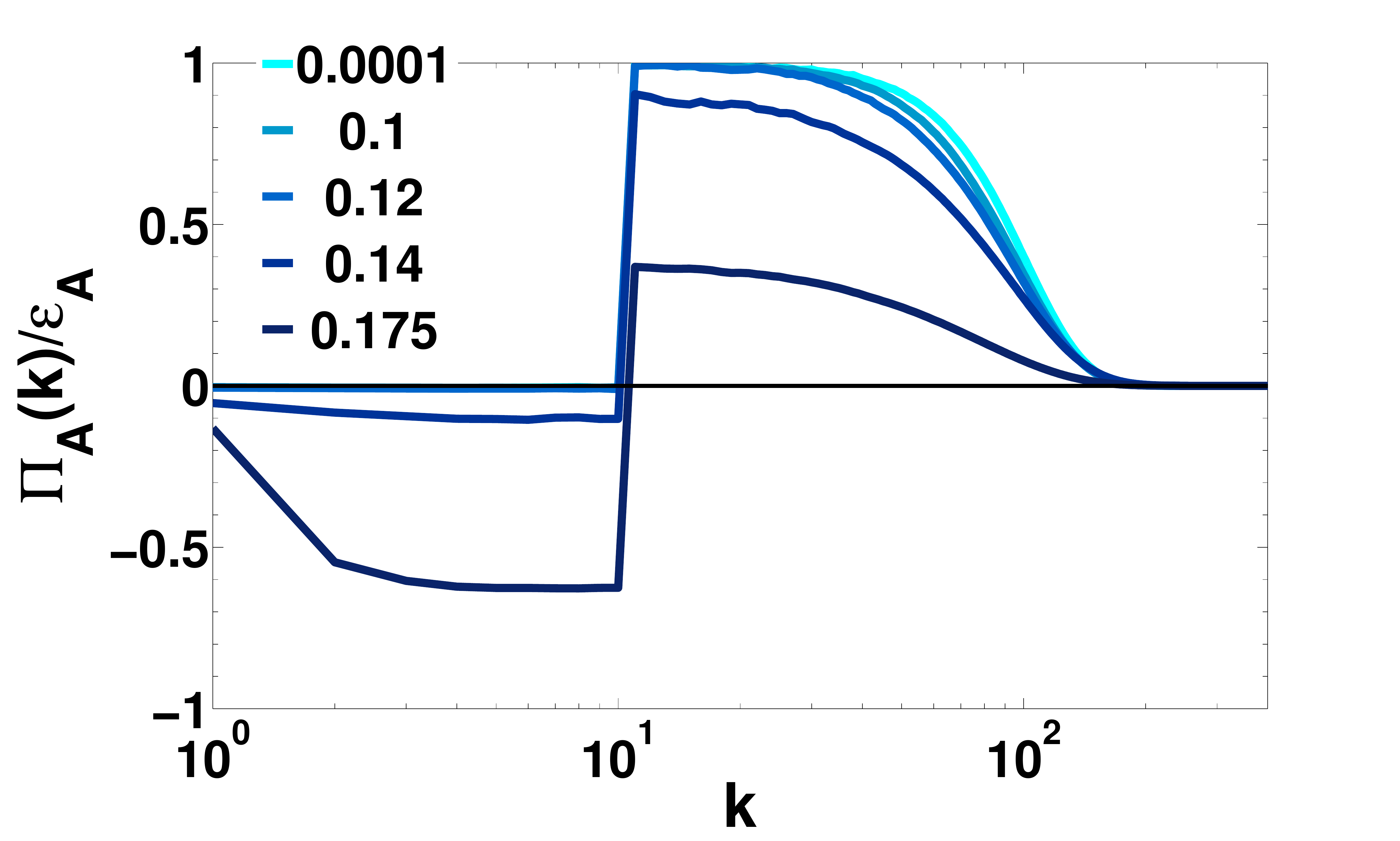}
\includegraphics[scale=0.15]{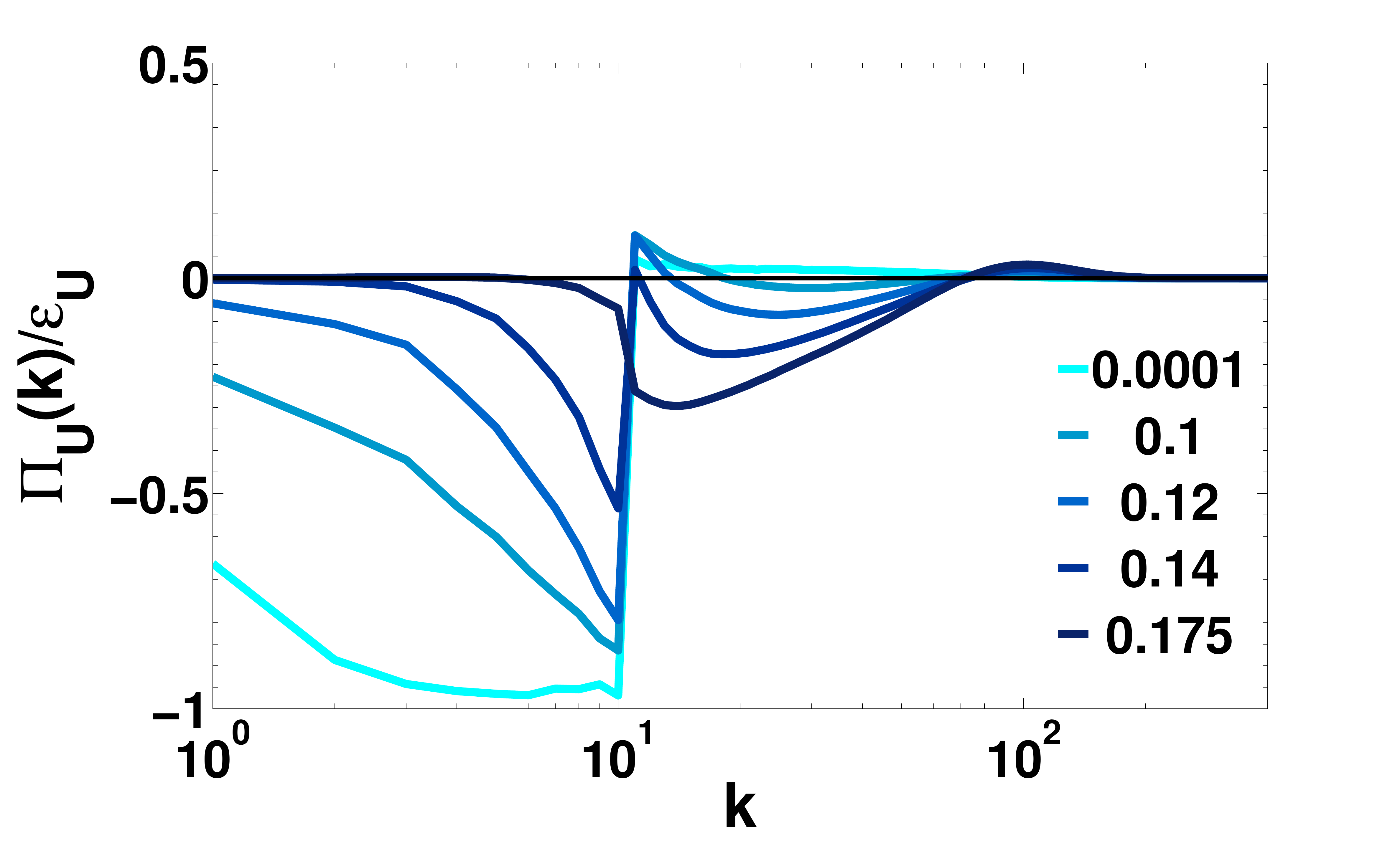}
\includegraphics[scale=0.15]{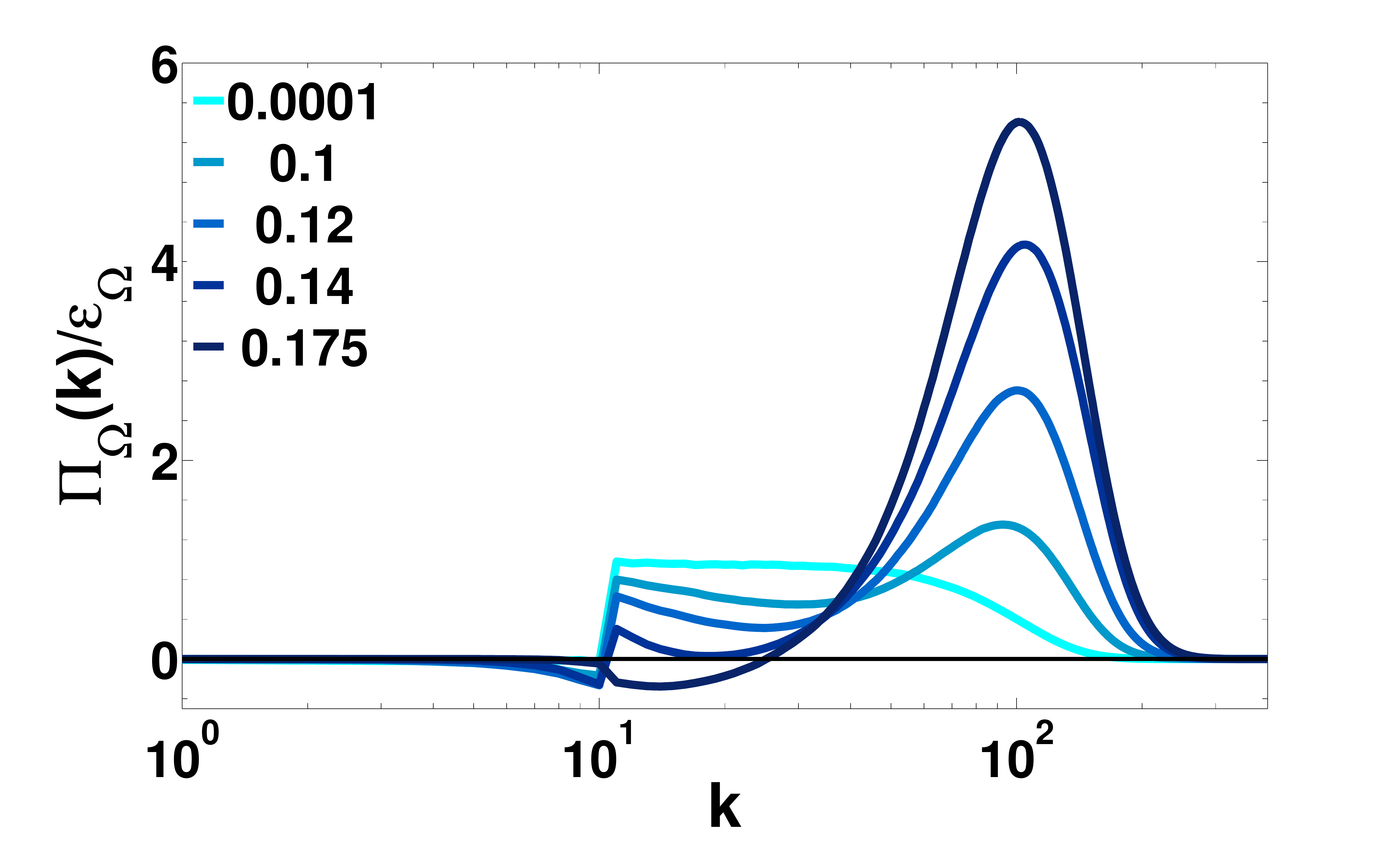}
\caption{\label{fig:Fluxes} The four plots from top are normalized fluxes of total energy $\Pi_{_E}/\epsilon_{_E}$, square vector potential $\Pi_{_A}/\epsilon_{_A}$, kinetic energy $\Pi_{_U}/\epsilon_{_U}$, and the enstrophy $\Pi_{_\Omega}/\epsilon_{_\Omega}$ as functions of wave-number $k$ for different values of the control parameter $\mu_f$. Darker shades denote larger values of $\mu_f$ as mentioned in the legend. }
\end{figure}
The flux of a conserved quantity across scales gives the rate that the nonlinearity transports this conserved quantity across scale space. Figure \ref{fig:Fluxes} shows the four fluxes $\Pi_{_U}, \Pi_{_\Omega}, \Pi_{_E}, \Pi_{_A}$ as a function of the wave-number $k$ for different control parameter $\mu_f$ for the $k_fL=8$ case, corresponds to case B3 in table \ref{Runs2}. 
We start with the case $\mu_f= 0.0001 \ll 1$, where the vector potential acts like a passive scalar. 
An inverse flux of kinetic energy $\Pi_{_U}<0$ is observed at wave-numbers smaller than $k_f$
and a forward flux of enstrophy $\Pi_{_\Omega}>0$ for $k>k_f$. 
For $k>k_f$ the flux of square vector potential $\Pi_{_A}>0$ has a forward cascade as expected for passive scalars. 
Since the magnetic field is small the flux of total energy is equal to the flux of the kinetic energy with a correction of the order $\mu_f^2$, $\Pi_{_E} = \Pi_{_U} + O(\mu_f^2)$.  
\begin{figure}
\includegraphics[scale=0.15]{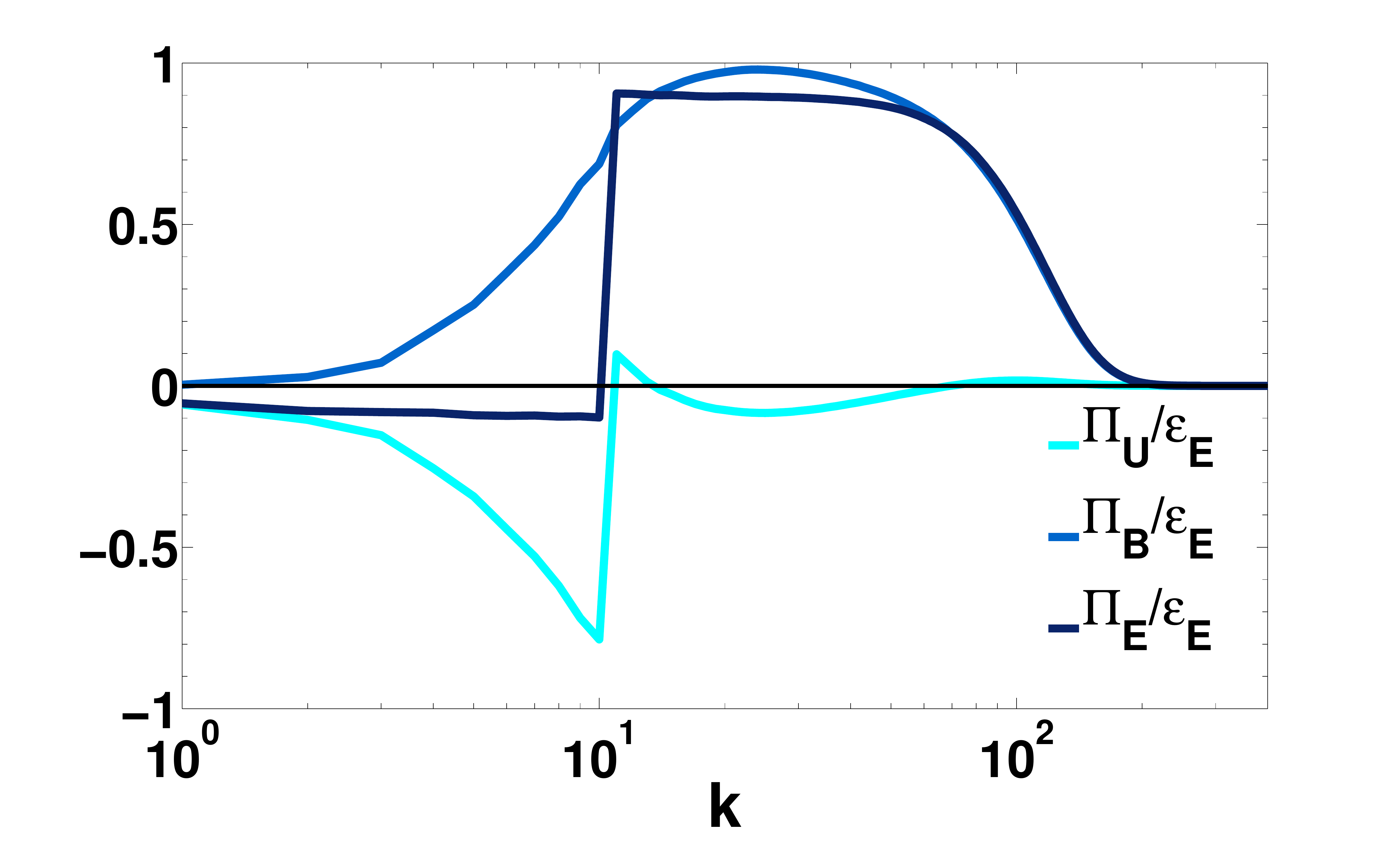}
\caption{\label{fig:lorentzflux} The normalized flux of $\Pi_{_U}, \Pi_{_B}, \Pi_{_E}$ for the control parameter $\mu_f \lesssim \mu_c$ is shown as a function of $k$, with parameters corresponding to case B3 in table \ref{Runs2}.}
\end{figure}
For larger values of $\mu_f$ the conservation of ${\Omega}$ breaks down progressively and the vector potential stops acting like a passive scalar. The Lorentz force injects enstrophy into the flow leading to non constant flux of $\Pi_{_\Omega}, \Pi_{_U}$. 
The enstrophy flux is modified most strongly at the largest wave-numbers where 
the curl of the Lorentz force $F_L = {\bf \hat{e}_z} j \times{\bf  b}$ is larger. 
For values close to the critical point $\mu_f \lesssim \mu_c$ we have a simultaneous forward and inverse cascade of total energy. 
As $\mu_f $ becomes larger than $ \mu_c$ the cascade of $\Pi_{_E}$ becomes strictly forward and an inverse cascade of $\Pi_{_A}$ is established.  

To further demonstrate the two processes involved in transferring energy to and from the large scales we calculate $\Pi_{_B} = \Pi_{_E} - \Pi_{_U}$. 
$\Pi_{_U}$ expresses the energy flux from terms that do not involve the magnetic field, while $\Pi_{_B}$ is the energy flux from terms that involves the magnetic field.
Figure \ref{fig:lorentzflux} shows $\Pi_{_U}, \Pi_{_B}, \Pi_{_E}$ as a function of $k$ for a $\mu_f$ which is close but smaller than the critical point. 
The two components of the fluxes $\Pi_{_U}$ and $\Pi_{_B}$ are not constant neither in the forward nor in the inverse cascade.
$\Pi_{_U}$ is negative (inverse) at almost all wave-numbers while  $\Pi_{_B}$ is positive (forward) at all wave-numbers.
Thus energy is transported by the velocity field from the small scales to the large scales,
while the opposite role is played by the magnetic field that transports energy from the large scales to the small scales.
The weak but constant energy flux observed at large scales is then the result of a counter balance of the  inverse flux of the kinetic energy
with the forward cascade of magnetic energy. This in part also explains the large fluctuations observed in the flux of energy in \cite{Kanna14}.

\subsection{Spectras} 

In the previous work \cite{Kanna14}, the spectra of kinetic and magnetic energy $E_u (k), E_b (k)$ were shown near the transition
focusing on the behavior of the large scales. 
For values of $\mu_f$ smaller than the critical value $\mu_{_Ec}$, 
the inverse flux of kinetic energy formed spectra proportional to $k^{-5/3}$ at large scales for the velocity field. 
The magnetic field showed equipartition of $A$ among large scales which led to the spectra $E_b(k) \sim k^{+3}$. 
For values of $\mu_f$ larger than $\mu_{_Ac}$ the magnetic energy starts to form an exponent of $k^{-1/3}$ due to the inverse cascade of $A$ \cite{Pandit, Biskamp03}.

\begin{figure}[!htb]
\includegraphics[scale=0.2]{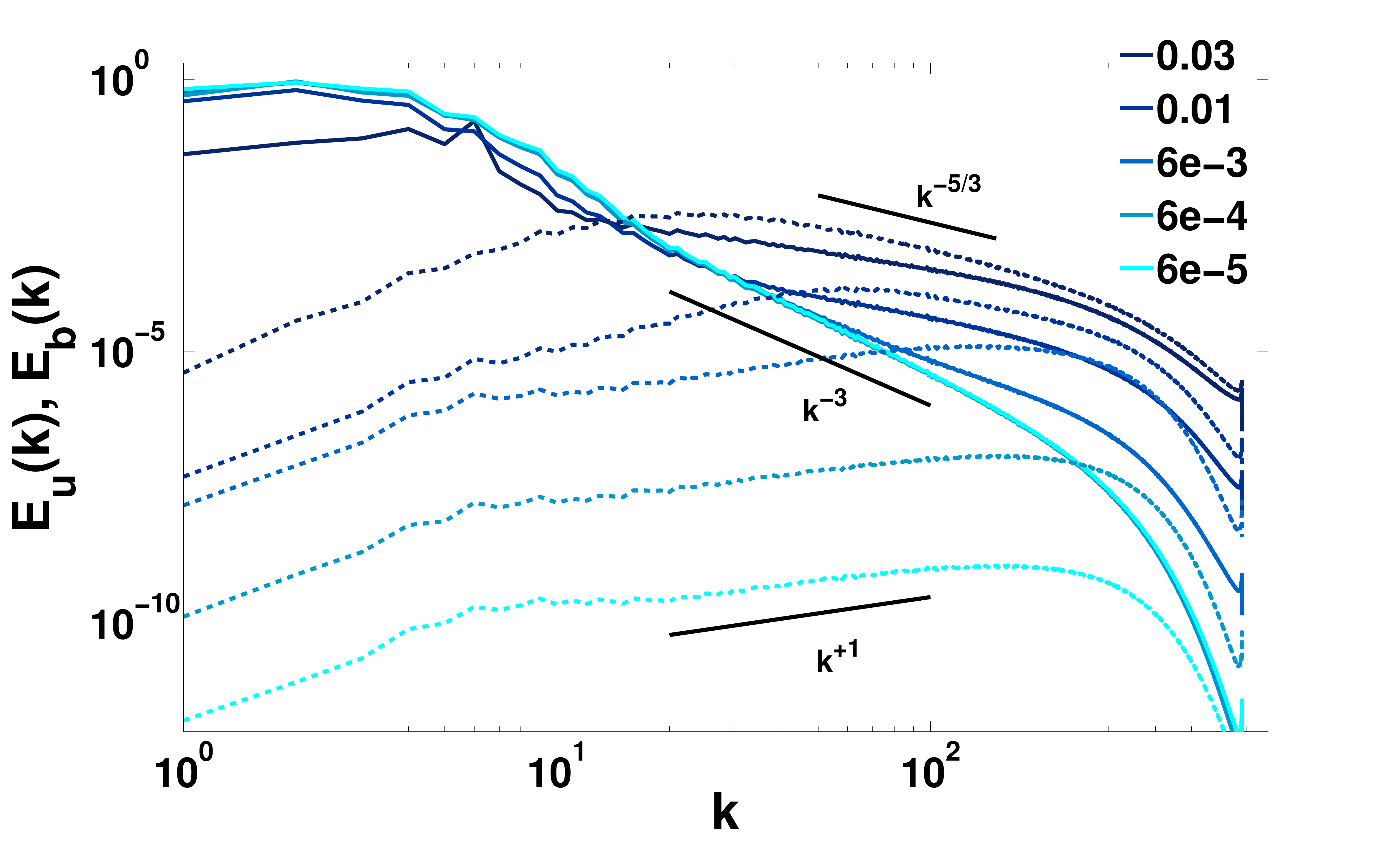}
\caption{\label{fig:spect} The spectra of the kinetic $E_u$ and the magnetic energy $E_b$ as functions of $k$ are shown for different value of $\mu_f$. Dotted line represents the magnetic energy spectra and the continuous line represents the kinetic energy spectra with darker shades corresponding to larger $\mu_f$. Black lines represent the possible exponents of the spectra. }
\end{figure}

Here we focus on the spectra at wave-numbers larger than $k_f$ and how they change as the parameter $\mu_f$ is increased
from infinitesimal values.  
%
In order to maximize the inertial range for the forward cascade we performed runs with $k_f=4$, with $Re^{+}=1.4\,10^7,Re^{-}=1.8\,10^3$. 
The kinetic energy $E_u(k)$ and the magnetic energy $E_b(k)$ spectra are shown in figure \ref{fig:spect} for few values of the control
parameter $\mu_f$. 
For these set of runs the parameter $\mu_f$ was kept smaller than its critical value $\mu_{_Ec}$ as the spectra in the forward
cascade did not change significantly for $\mu_f>\mu_{_Ec}$. 

\begin{figure}[!htb]
\includegraphics[scale=0.175]{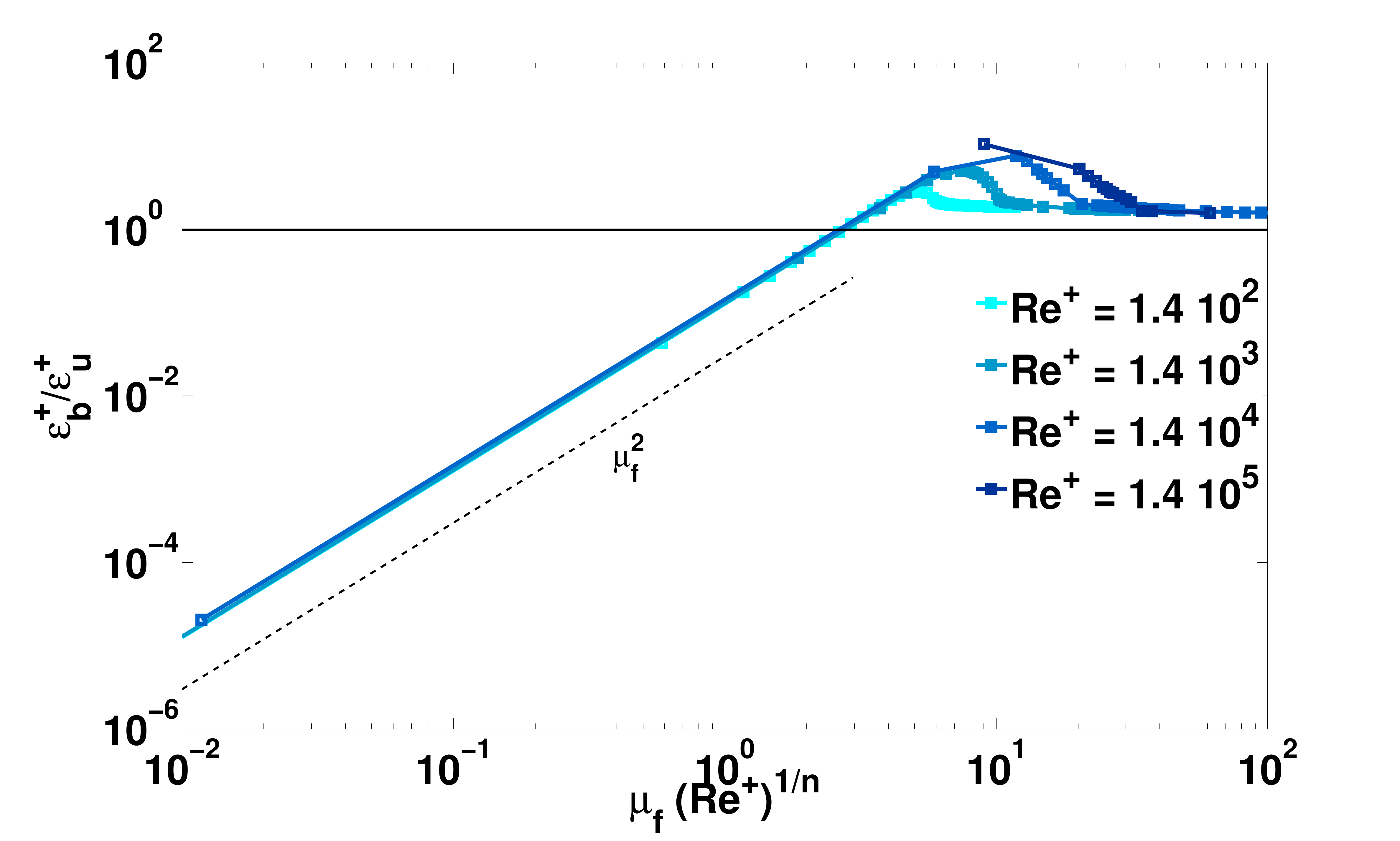}
\caption{\label{fig:epsart} The ratio of the dissipation at small scales of the magnetic and the kinetic field $\epsilon_b^+/\epsilon_u^+$ are shown as function of the rescaled control parameter $\mu_f \, \left( Re^+ \right)^{1/n} $. Four different value of $Re^+$ are considered as mentioned in the legend, for runs in table \ref{Runs2}. The continuous black line represents the value $\epsilon_b^+/\epsilon_u^+ = 1$. The dotted black line shows the scaling at the passive scalar limit. Darker shades correspond to larger $Re^+$.}
\end{figure}

For the lowest value of $\mu_f$ the magnetic energy spectrum is smaller than the kinetic energy spectrum at all scales
and  corresponds to the passive scalar limit. The forward cascade of enstrophy leads to a kinetic energy spectrum 
$E_u \propto \Pi_{_\Omega}^{2/3} k^{-3}$.
The magnetic field on the other hand is proportional to the gradients of a passively advected scalar, forms 
a spectrum $E_b \propto \Pi_{_A}\Pi_{_\Omega}^{-1/3} k^{+1}$.
%
%
However since the magnetic spectrum has a larger exponent than the kinetic energy spectrum the two spectra will meet
at a wave-number $k_m$ that can be obtained by equating the two relations for $E_u$ and $E_b$ and leads to
\begin{equation}
k_m \propto (\Pi_{_\Omega}/\Pi_{_A})^{1/4}\propto \mu_f^{-1/2} k_f.
\label{km}
\end{equation}
This however is realized provided that the inertial range, limited by the dissipation, is long enough so that $k_m \le k_\nu$.
This in turn implies that the transition from the passive scalar regime ($b_{\nu} \ll u_{\nu}$) to a regime at which magnetic and kinetic energy
are comparable at the small scales ($b_{\nu} \sim u_{\nu}$) occurs for a value $\mu_f$ that we will refer to as $\mu_{_{NL}}$. To have an estimate for $\mu_{_{NL}}$ we look at the forward dissipation length scale in the passive limit given by $\ell_{\nu} = k_{\nu}^{-1} \propto k_f^{-1} \left( Re^+ \right)^{-1/2n}$. 
By equating $k_m$ with $k_\nu$ we have the scaling, 
\begin{eqnarray} 
\mu_{_{NL}} \propto \left( \frac{k_f}{k_{\nu}} \right)^{2} \sim \left( Re^{+} \right)^{-1/n}
\end{eqnarray}
This scaling is clearly demonstrated in figure \ref{fig:epsart} where the ratio $\epsilon_b^+/\epsilon_u^+$ is plotted
as a function of $\mu_f \;  (Re^+)^{1/n}$ that collapses all curves together for different values of $Re^+$ from Table \ref{Runs2}. 
The ratio $\epsilon_b^+/\epsilon_u^+$ expresses the ratio of magnetic to kinetic energy
weighted at the small scales. Clearly for all runs $\epsilon_b^+/\epsilon_u^+$ is smaller than unity 
for all values of $\mu_f$ such that $\mu_f (Re^+)^{1/n}\lesssim 3$ that marks the beginning of the nonlinear feed back
of the magnetic field. Note that the onset of the nonlinear behavior $\mu_{_{NL}}$ is different from the 
the critical value $\mu_{c}\propto (Re^+)^{-1/2n}$ that marks the end of the inverse cascade.

For values of $\mu_f$ larger than $\mu_{_{NL}}$ the passively advected magnetic field spectra appears to still 
hold but only for wave-numbers for which $k<k_m$. This can be seen in figure \ref{fig:spect} for the kinetic and magnetic spectra
in the range of wave numbers for which $E_b\ll E_u$. For larger wave numbers the observed spectral slopes change.
For sufficiently large $\mu_f$ and for values $k>k_m$ there is a new power law behavior observed that is close to
 $k^{-5/3}$ for both $E_b$ and $E_u$ spectra. 
This is in agreement with the 2D MHD prediction of a constant forward energy flux to small scales. Thus the initial power laws 
$E_b(k<k_m) \sim k^{+1}, E_u(k<k_m) \sim k^{-3}$ transitions to $E_b(k>k_m) \sim k^{-5/3}, E_u(k>k_m) \sim k^{-5/3}$ at a fixed value of $\mu_f$. As we increase the value of $\mu_f$ the magnetic energy becomes larger thus the point $k_m$ 
moves closer to the forcing length scale $k_f$ as expected from the estimate in equation \ref{km}. 
These arguments finally break down when $\mu_f$ comes close to the critical value $\mu_{_{E}c}$.
%

\section{Spatial behavior \label{SP}} 

\subsection{Structures} 
\begin{figure}[!htb]
\includegraphics[scale=0.2]{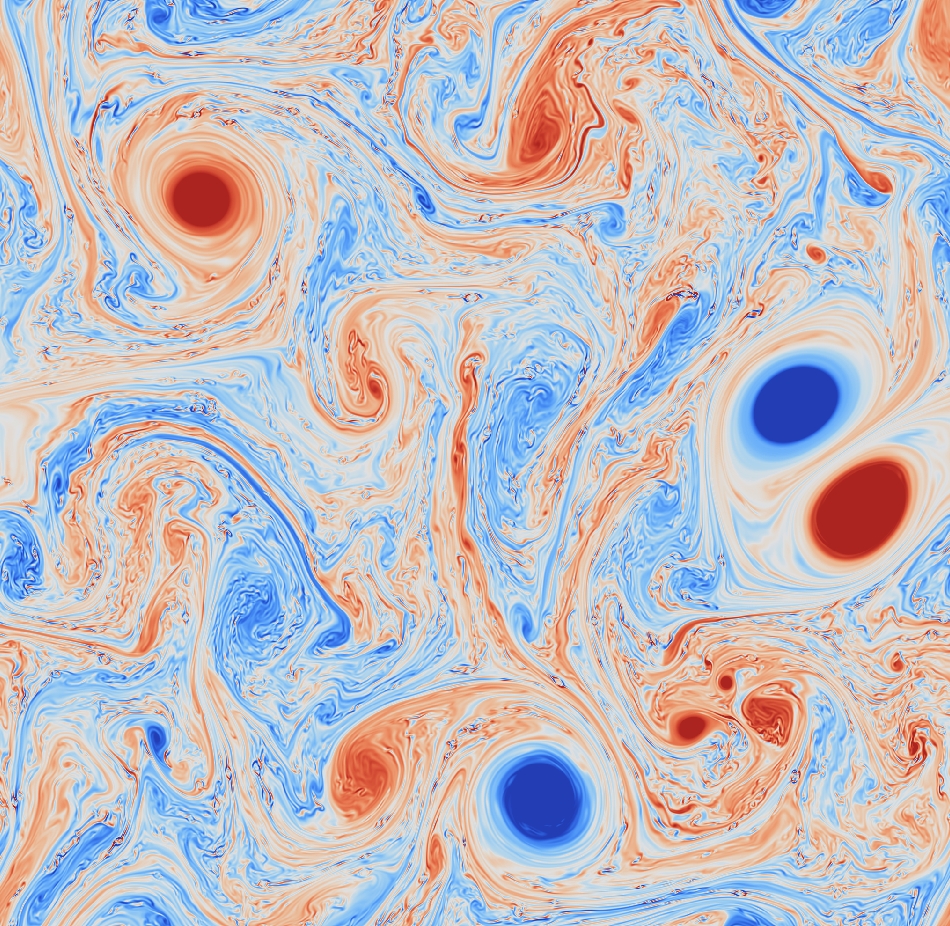}
\includegraphics[scale=0.2]{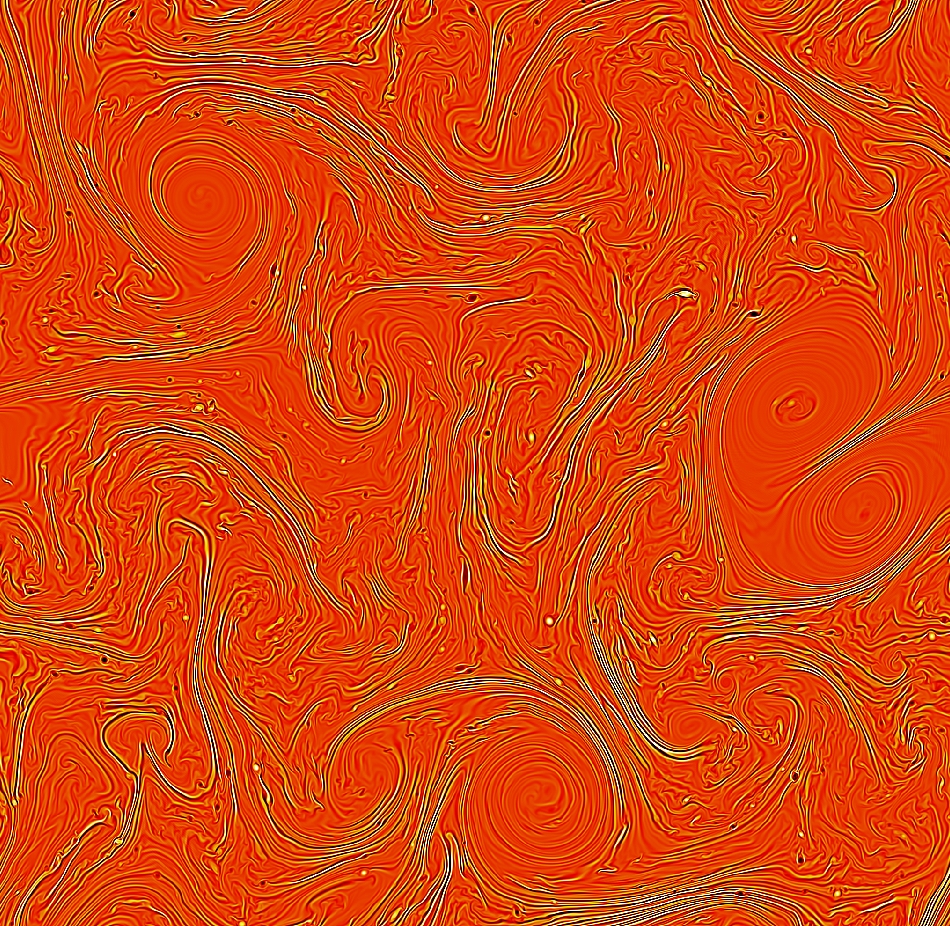}
\caption{\label{fig:contour} Contour plots of the vorticity $\omega$ on top and the current $j$ in bottom are shown for a value $\mu_{_{NL}} < \mu_f < \mu_c$ with $k_f \, L = 4$. Both large scale and small scales structures are found due to the dual cascade of energy and square vector potential. }
\end{figure}

%
The different processes and the different phases discussed in the previous sections 
have a direct impact on the structures formed. 
For values of $\mu_f<\mu_{_{NL}}$ the flow forms large scale vortexes due to the inverse cascade of energy and weak small scale
vortex filaments due to the forward cascade of enstrophy. Properties of these structures have been extensively discussed in the
literature \cite{chen2006physical}. At the same time the vector potential, that acts as a passive scalar, forms also filamentary structures \cite{batchelor1959small, ottino1989kinematics}.
This behavior starts to change as $\mu_f$ becomes larger than $\mu_{_{NL}}$.  
Figure \ref{fig:contour} shows the vorticity $\omega$ and the current $j$ for $\mu_{_{NL}}<\mu_f < \mu_c$ and $k_fL=4$. 
The large scales structures seen in $\omega$ is due to the strong inverse cascade of the kinetic energy similar to $2D$ HD systems.
The small scale structures in $\omega$ field however are not similar to the ones of 2D HD turbulence 
but closer to $2D$ MHD systems. The current density in the same figure leads to the same conclusion. 
The current is concentrated in thin filaments aligned with the large scale shear
as for the case of an advected passive scalar. However these filaments are not `straight elongated structures' as the case 
of the gradients of a passive scalar but rather the filaments `wiggle' varying along the direction of the shear,
probably due to the appearance of nonlinear MHD (Alfven) waves. 

\begin{figure}[!htb]
\includegraphics[scale=0.2]{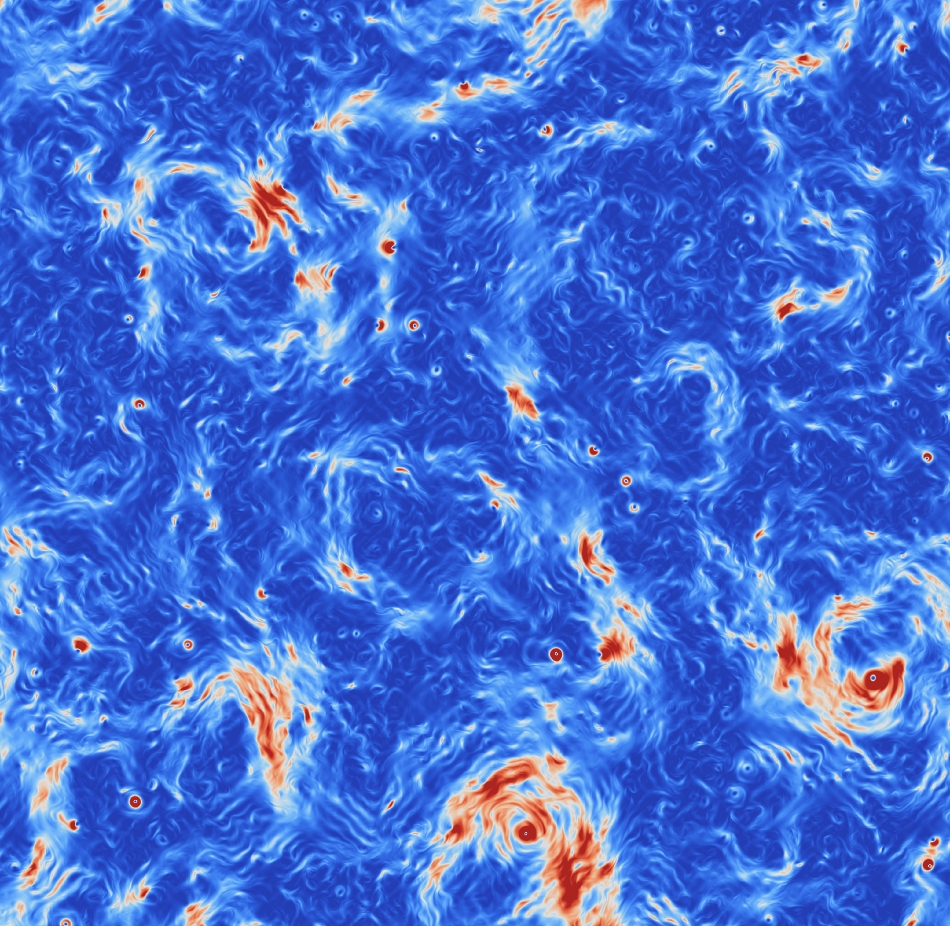}
\includegraphics[scale=0.2]{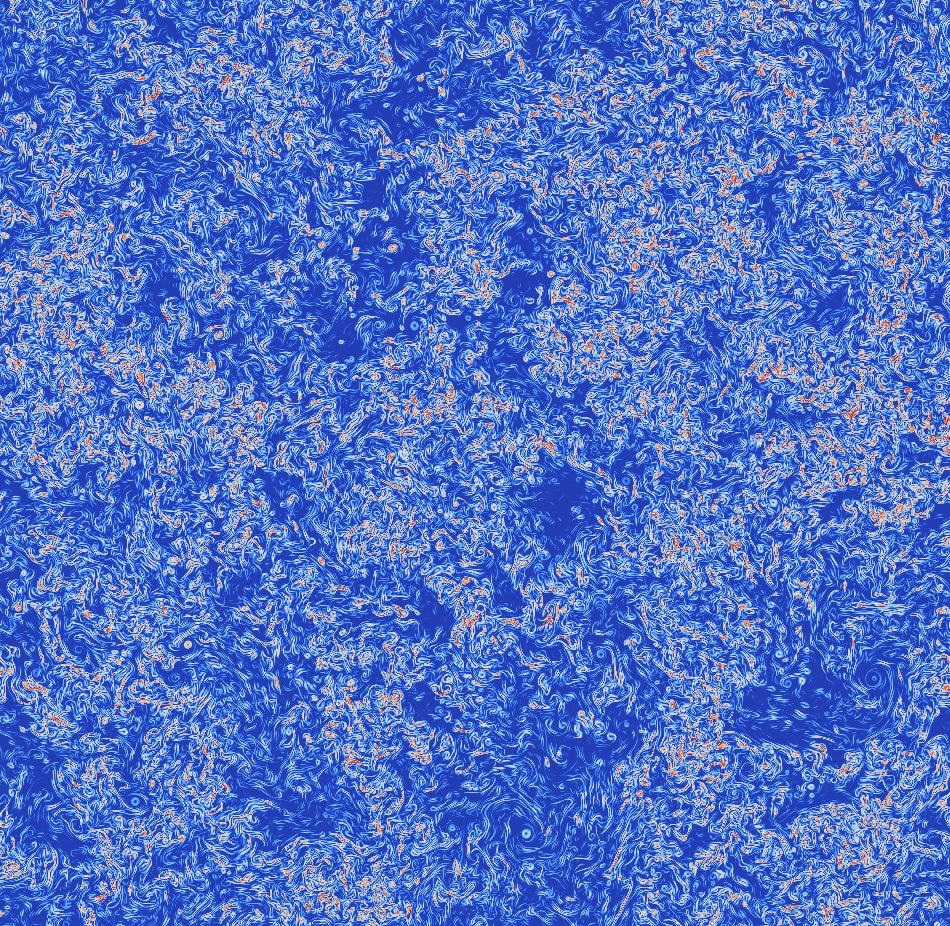}
\caption{\label{fig:cont2} Contour plots of the kinetic energy field on top and the magnetic energy field in bottom are shown for a value $\mu_f \lesssim \mu_{_Ec}$ with $k_f \, L = 64$. }
\end{figure}

Figure \ref{fig:cont2} shows the kinetic and the magnetic energy $E_u, E_b$ for a value $\mu_f \lesssim \mu_{_Ec}$ and $k_fL=64$.
A large portion of kinetic energy is in small scales with the existence however of scales 
much larger than the forcing scale, a clear evidence of an inverse cascade. 
The large scale structures in the kinetic energy are not space filling suggesting that the inverse cascade is weaker than 
the pure 2D HD. Magnetic energy is concentrated only in the small scales. 
However we note that there is an anti-correlation between the kinetic field and the magnetic field:
Regions of large velocity structures are correlated with regions of weak magnetic field and vice versa. 
Thus it appears that some      regions of the flow act as 2D HD  cascading energy inversely and 
                     other regions             act as 2D MHD cascading energy to the small scales 
                     in accordance with the process suggested in figure \ref{fig:lorentzflux}.   

\begin{figure}[!htb]
\includegraphics[scale=0.2]{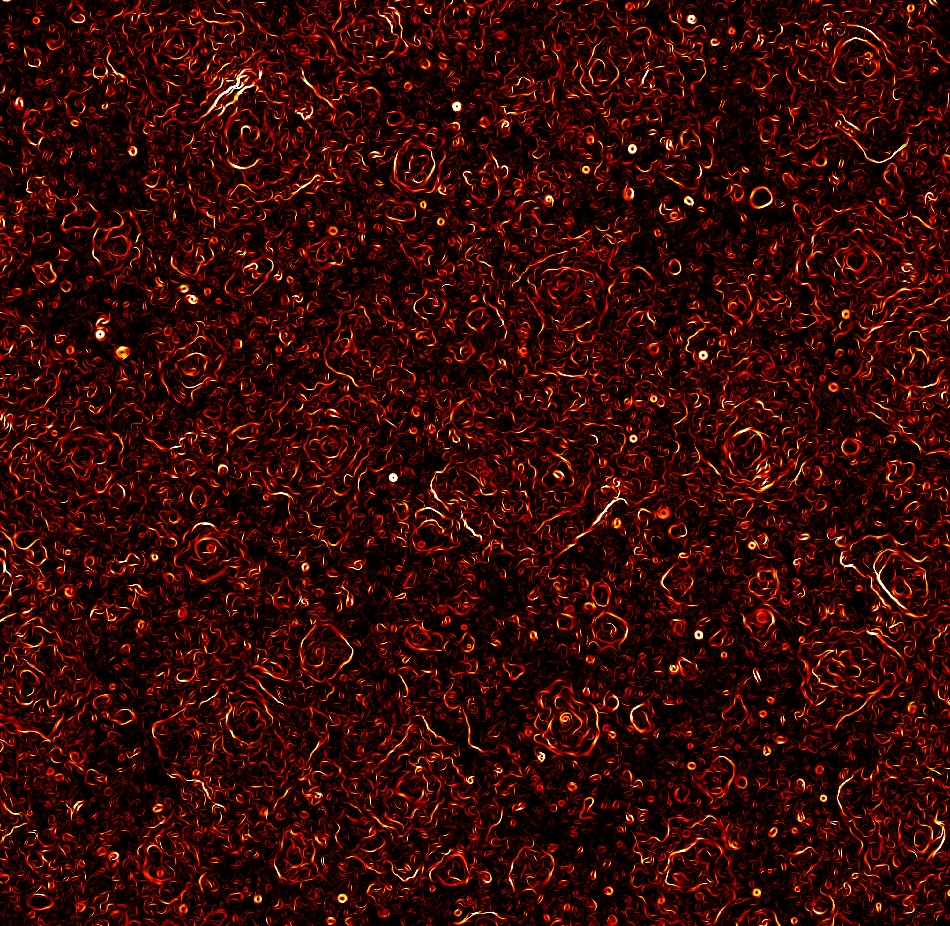}
\includegraphics[scale=0.2]{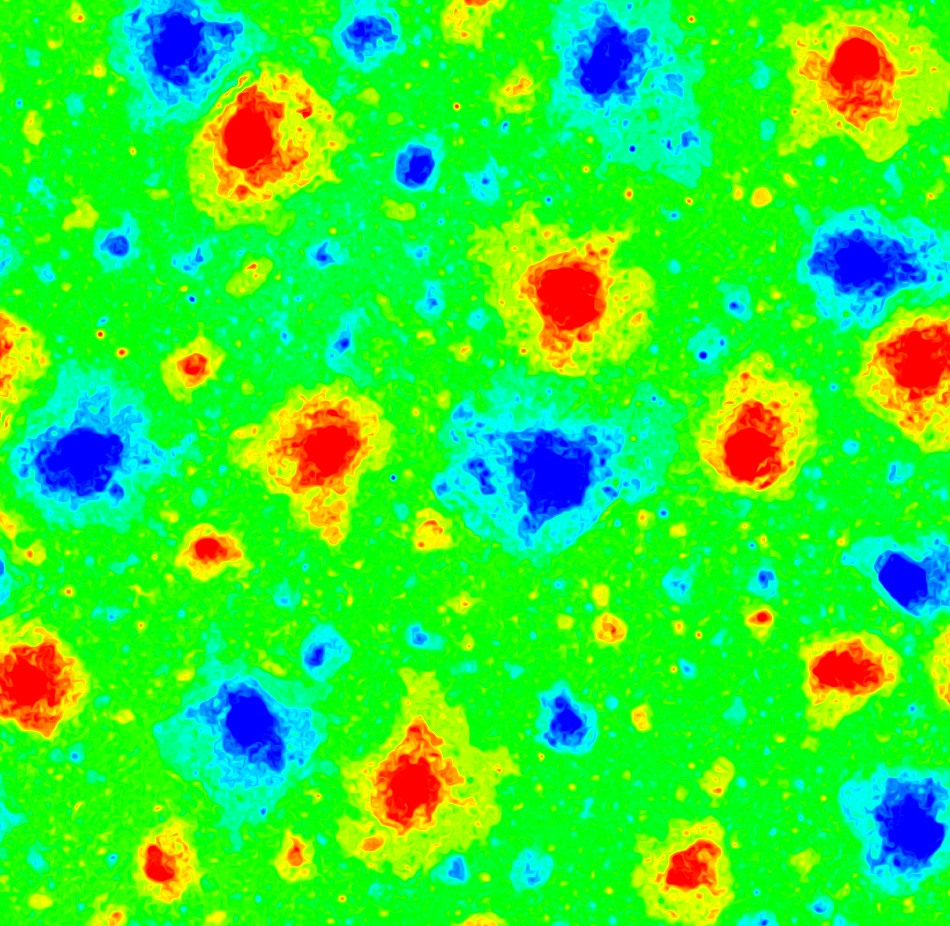}
\caption{\label{fig:cont3} Contour plots of the magnetic energy field on top and the vector potential in bottom are shown for a value $\mu_f \gtrsim \mu_{_Ac}$ with $k_f \, L = 64$. }
\end{figure}

Finally figure \ref{fig:cont3} shows the magnetic energy $E_b$ and the square vector potential for a value $\mu_f \gtrsim \mu_{_Ac}$
and $k_fL=64$. The kinetic energy (not shown) does not show any large scale structure. The magnetic energy is clearly small scale 
with a filamentary structure. However the current filaments are arranged so that the vector potential forms large scale
islands.  

\subsection{Spatial statistics} 

To be able to quantify the qualitative description given in the previous section 
we calculate distribution of differences of velocity $\delta u_{_L} \left( {\bf x, r}, t\right)$
and vector potential $\delta a \left( {\bf x, r}, t\right)$ between two points ${\bf x}$ and ${\bf x+r}$. Here subscript $L$ denotes longitudinal component of the vector $\delta u_{_L} = \delta {\bf u}_{_L} \cdot \hat{r}$.
Studies of 2D HD \cite{boffetta2012two, boffetta2000inverse, paret1998intermittency} have suggested that the inverse cascade in $2D$ turbulence is self-similar
(and possibly conformal invariant \cite{Bernard06}) in the sense that the pdfs of velocity difference from two points at distance
$r$ (that lies in the inverse inertial range) take the form $\propto f(\delta u_{_L}/r^\alpha)$ for some exponent $\alpha$ . 

\begin{figure}[!htb]
\includegraphics[scale=0.2]{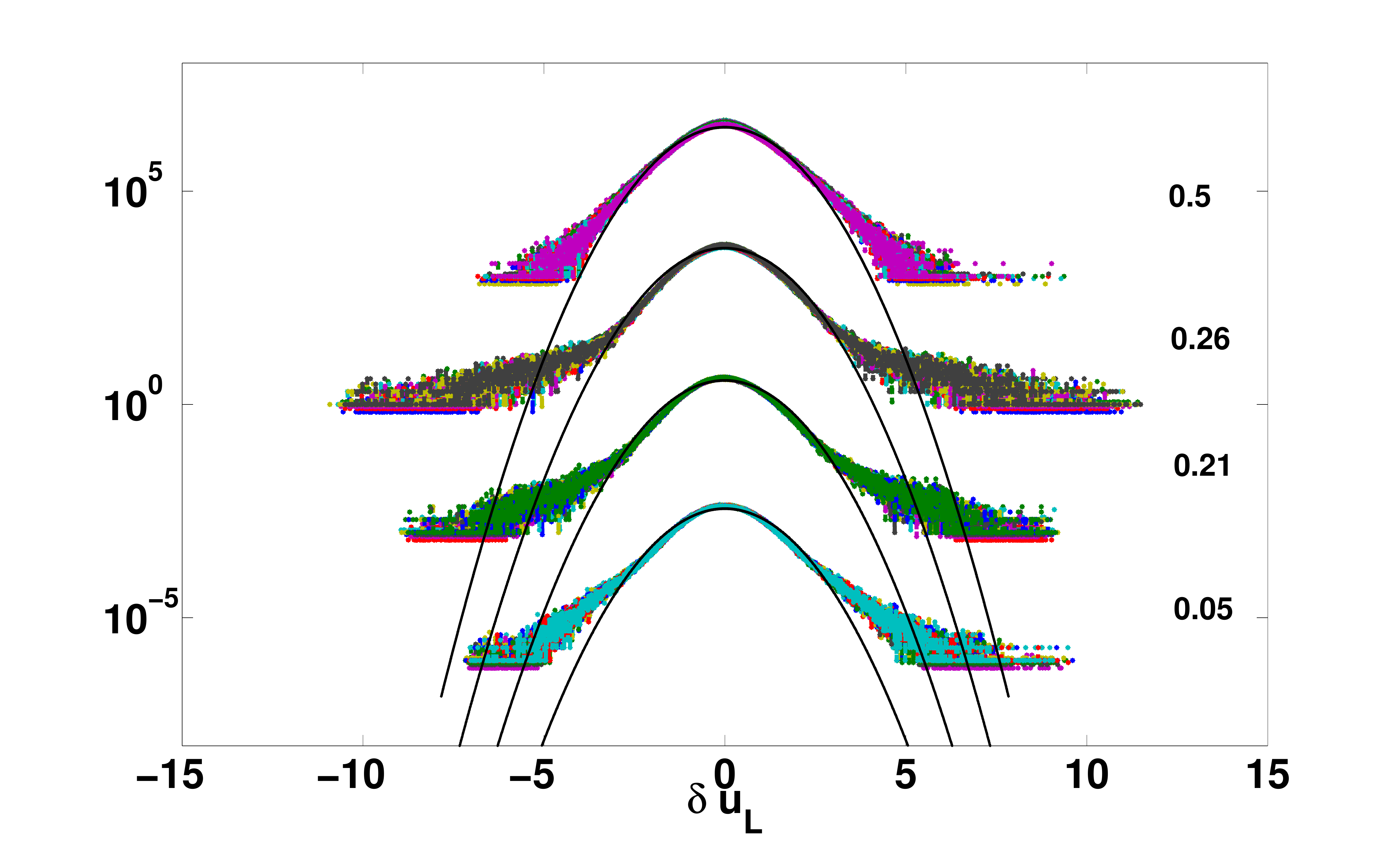}
\includegraphics[scale=0.2]{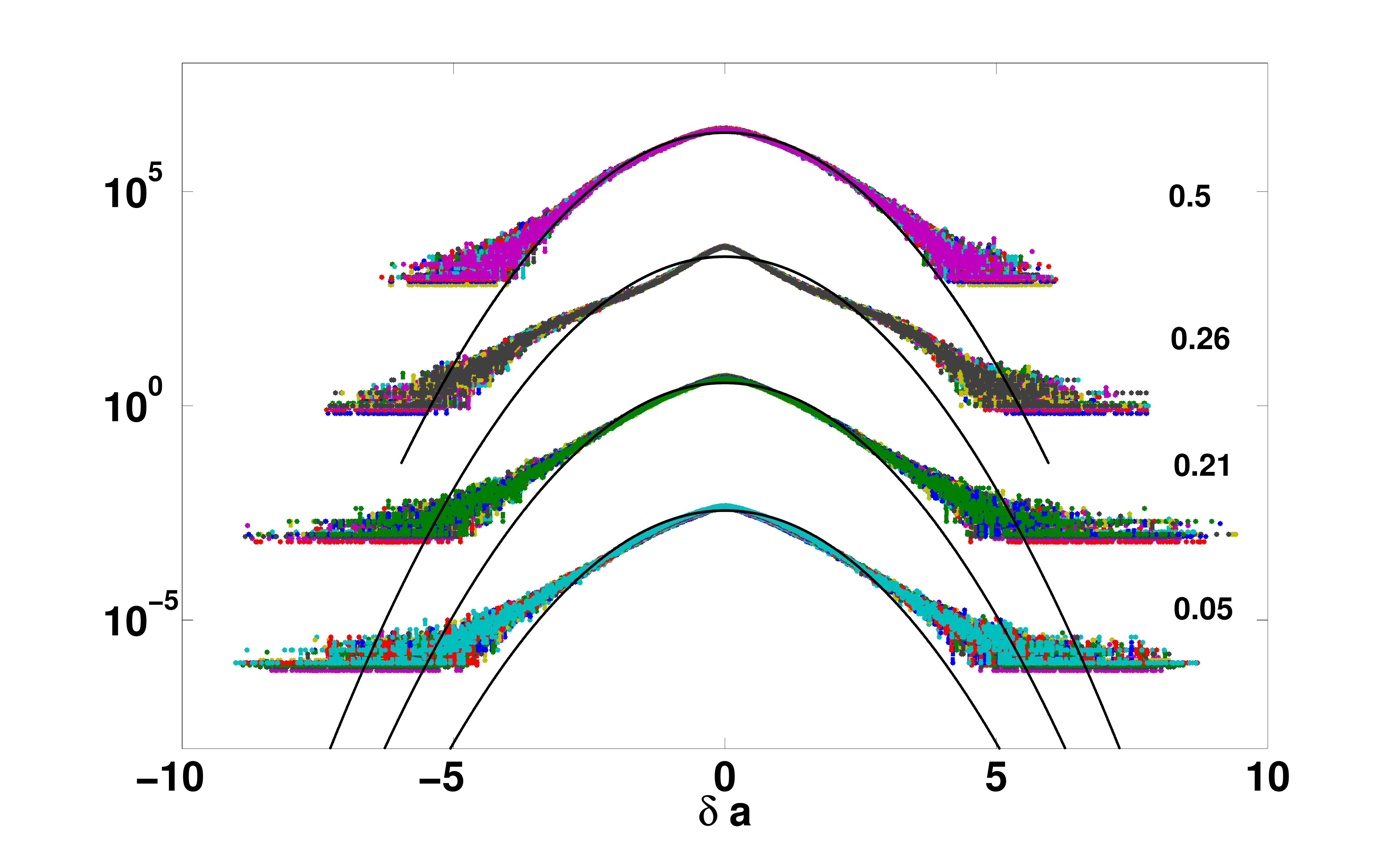}
\caption{\label{fig:pdfdiff} The normalized pdfs - $\delta u_{_L} \left( {\bf x, r}, t\right) / \left\langle \delta u_{_L}^2 \left( {\bf x, r}, t\right) \right\rangle^{1/2}$ and $\delta a \left( {\bf x, r}, t\right) / \left\langle \delta a^2 \left( {\bf x, r}, t\right) \right\rangle^{1/2}$ on top and bottom respectively are shown for different values of $\mu_f= 0.05, 0.21, 0.26, 0.5$ represented on the side. Different values of $\mu_f$ are shown by displacing the pdfs vertically. Different colors represent different values of $r$. Black lines represent the gaussian curve with unity variance. }
\end{figure}

The probability distribution function (pdf) of the normalized longitudinal velocity difference $\delta u_{_L} \left( {\bf x, r}, t\right) / \left\langle \delta u_{_L}^2 \left( {\bf x, r}, t\right) \right\rangle^{1/2}$ 
and the normalized vector potential difference $\delta a \left( {\bf x, r}, t\right) / \left\langle \delta a^2 \left( {\bf x, r}, t\right) \right\rangle^{1/2}$ 
for different values of $r$ 
are shown in figure \ref{fig:pdfdiff}. 
The pdfs were calculated using the results from the resolution runs in table \ref{Runs1}.
Different values of the control parameter $\mu_f$ are shown by successive vertical shifts in the $y$-axis. 
Different colors corresponding to different values of $r$ ranging from $r^{-1} = 0.9 \times k_f$ to $ 0.1 \times k_f$ 
corresponds to the inverse cascade range. 
Black curves correspond to a normalized Gaussian curve of unit variance. 
For all the values of $r$ the pdfs collapse on each other indicating self-similarity. 
This is seen to be consistent across transition as the value of $\mu_f$ varies from $0.05-0.5$ with 
$\mu_c \sim 0.23$ and $k_f \, L = 16$ see table \ref{Runs1}. Similar behavior is observed for larger values of $k_f \, L = 32, 64$. For pdf of $\delta u_{_{L}}$ close to the transition the presence of larger tails suggests higher probability of 
extreme events.

From the different moments that we can examine, of particular interest are the third order moments that are related
for homogeneous and isotropic flows by Karman-Howarth type relations for the total energy. 
The Karman-Howarth relation for $E$ is given by \cite{politano98}, 
\begin{eqnarray}
-\frac{3}{2} \epsilon_{_E} r = \left\langle \delta u_{_L}^3 \left( {\bf x, r}, t \right) \right\rangle - 6 \left\langle b_{_L}^2 \left( {\bf x}, t \right) \delta u_{_L} \left( {\bf x, r}, t \right) \right\rangle
\end{eqnarray}  

%
For the energy balance we have a competition between the inverse cascade of kinetic energy created by 
$\left\langle \delta u_{_L}^3 \left( {\bf x, r}, t \right) \right\rangle$ and 
the forward cascade created by $\left\langle \delta u_{_L} \left( {\bf x, r}, t \right) b_{_L}^2 \left( {\bf x, r}, t \right) \right\rangle$. 
Thus the direction of cascade is determined by which term dominates.
\begin{figure}[!htb]
\includegraphics[scale=0.2]{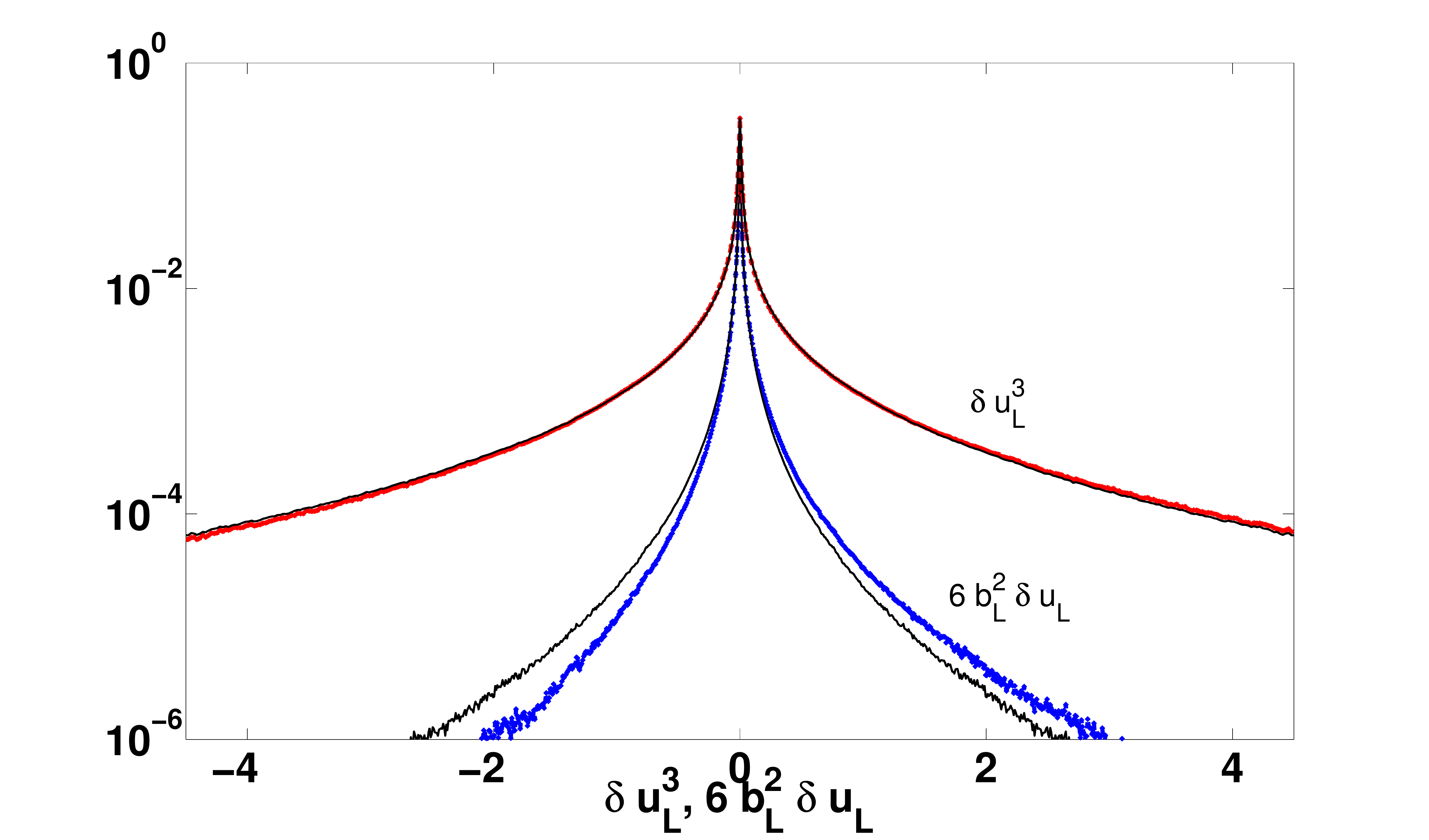}
\includegraphics[scale=0.2]{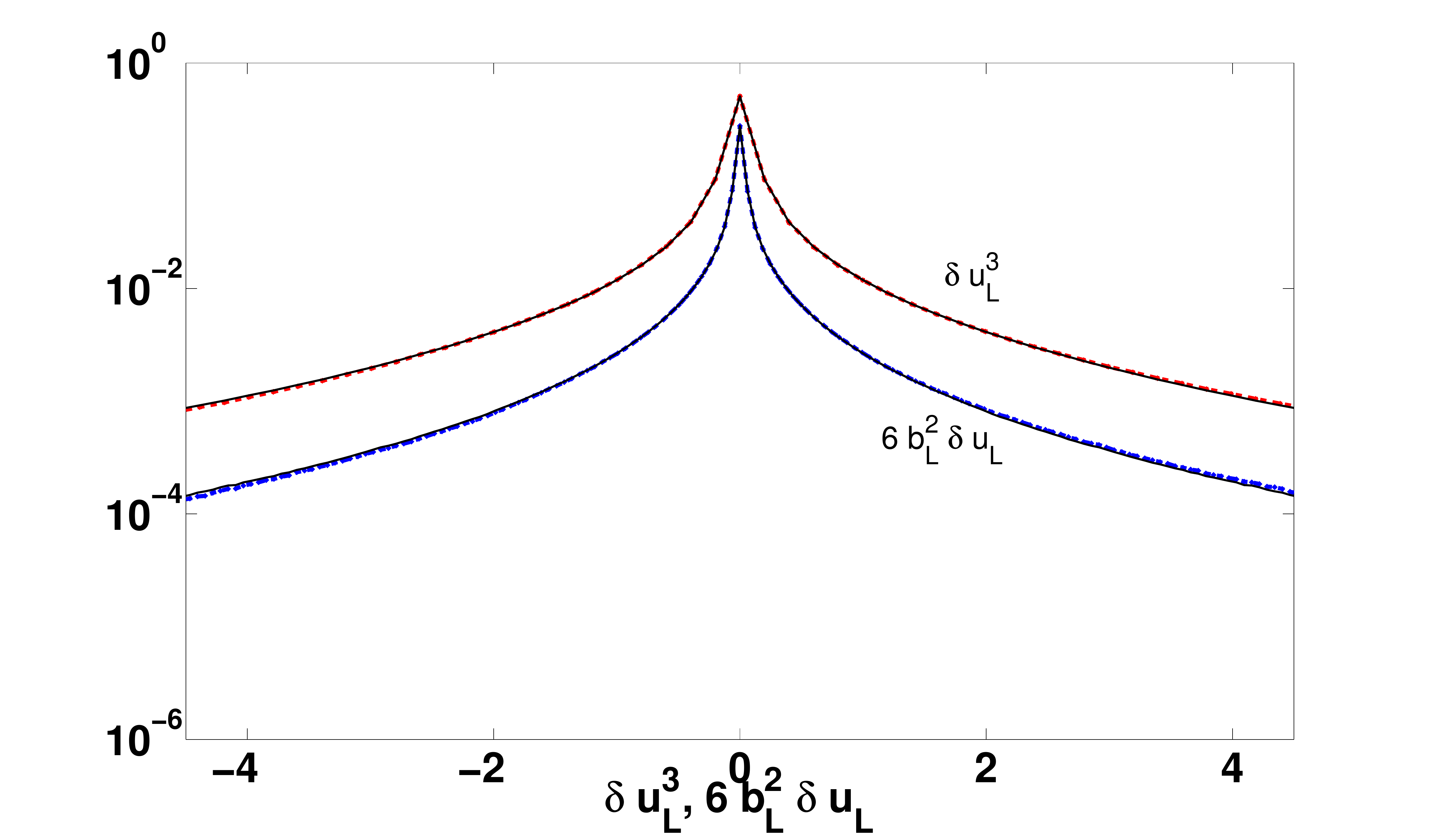}
\caption{\label{fig:pdfener} Pdfs $\delta u_{_L}^3$ in red and $6 b_{_L}^2 \delta u_{_L}$ in blue  are shown for two different values of $\mu_f$. $\mu_f=0.21 < \mu_c$ is the one on top and $\mu_f = 0.26 > \mu_c $ is the bottom one. Black lines denote the symmetric part of the pdfs.}
\end{figure}
%

The pdfs of $6 \left\langle b_{_L}^2 \left( {\bf x, r}, t\right) \, \delta u_{_L} \left( {\bf x, r}, t\right) \right\rangle $ 
and $\left\langle \delta u_{_L}^3 \left( {\bf x, r}, t\right) \right\rangle$ slightly before and after transition 
are shown in figure \ref{fig:pdfener}. The values of the control parameter are $\mu_f = 0.21$ and $0.26$. 
Self-similarity across all $r$ in the inverse cascade scales allows us to average across $r$ in the inverse cascade inertial range. 
The symmetric part of the pdf (where for a function $f(x)$ it is defined as $\left( f(x)+f(-x) \right)/2$) 
is shown by the black line. The asymmetry in the curve gives an indication for the direction of the cascade of the energy across scales. 
For $\mu_f < \mu_{_{E}c}$ (top panel) 
the pdfs have exponential asymmetric tails with the asymmetry in favour of the positive values.
%
Thus based on the Karman-Howarth relation the $\left\langle \delta u_{_L}^3 \left( {\bf x, r}, t\right) \right\rangle$ term
leads to a flux of energy towards the large scales and the term 
$-6 \left\langle b_{_L}^2 \left( {\bf x, r}, t\right) \, \delta u_{_L} \left( {\bf x, r}, t\right) \right\rangle $ 
leads to a flux of energy towards the small scales.
For values of $\mu_f$ larger than the critical point $\mu_f > \mu_{_{A}c}$ the tails remain non-Gaussian 
but the asymmetry in the pdfs of both terms is diminished, thus the exchange of energy with the large scales is suppressed.


\section{Summary and Perspectives } 


\begin{figure}[!htb]
\includegraphics[scale=0.3]{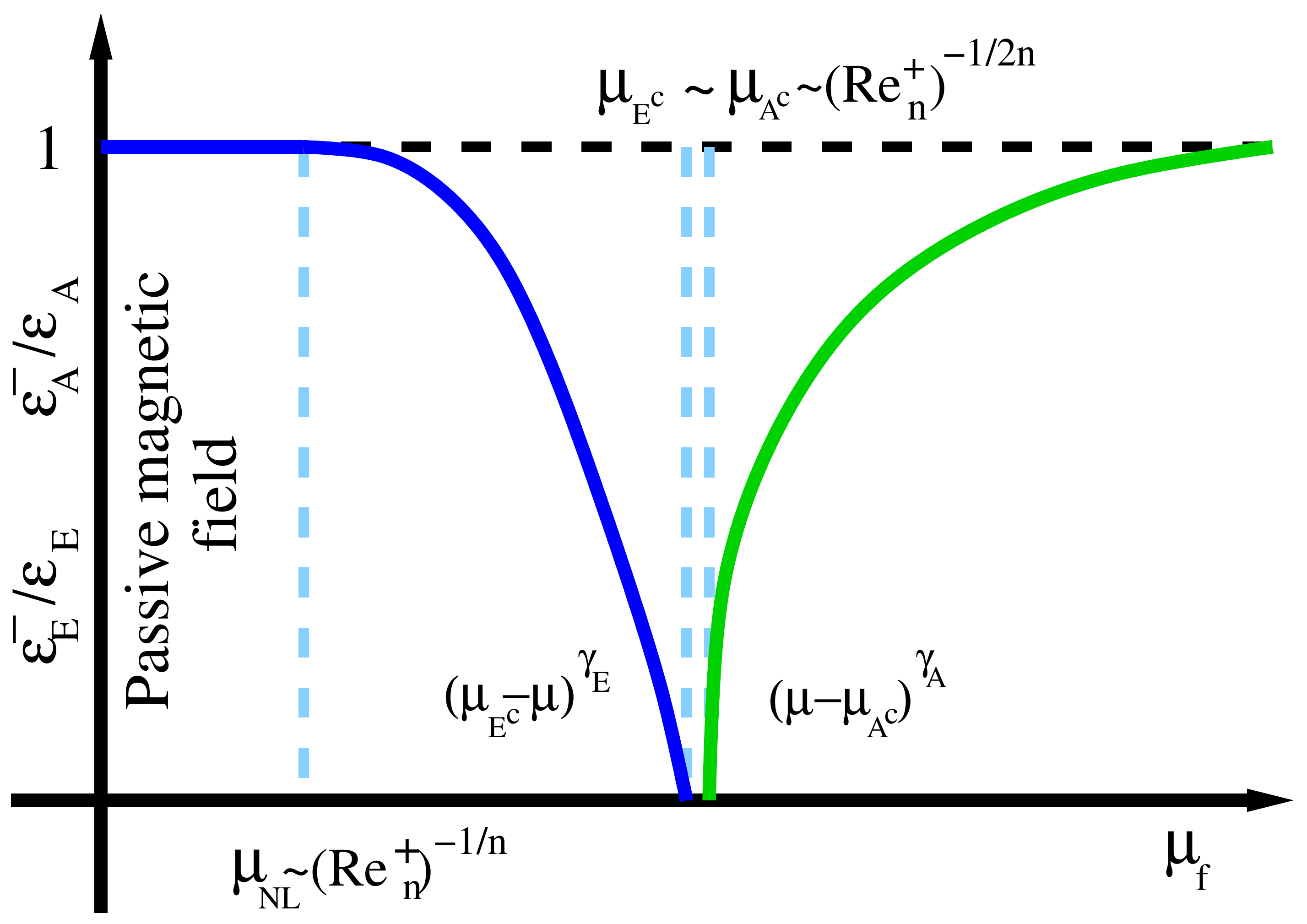}
\includegraphics[scale=0.3]{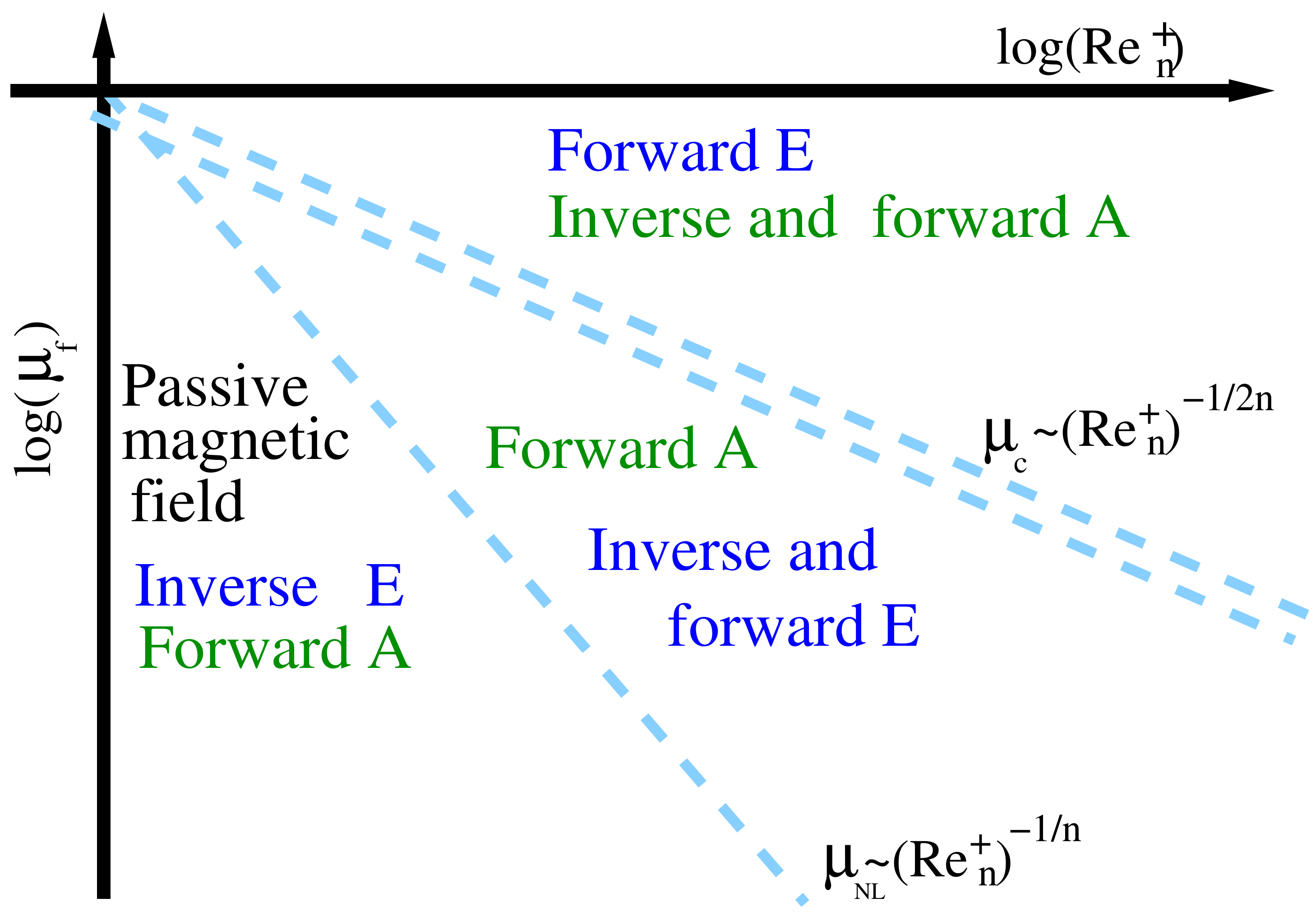}
\caption{\label{fig:Conclusion}  Top panel: the variation $\epsilon_{_E}^-/\epsilon_{_E}, \epsilon_{_A}^-/\epsilon_{_A}$ as a function of $\mu_f$. 
         Denoted are the critical points $\mu_{_Ec},\mu_{_Ac}$ and the region where the nonlinear behaviour starts $\mu_{_{NL}}$. $Re_n^{\pm}$ denoting the Reynolds number calculated based on the order of laplacian $n$.
         Lower panel: The phase space diagram in the parameter space $Re^{+}$, $\mu_f$.  $Re^-, k_f \, L$ is assumed to be large or infinite. }
\end{figure}

We have studied by an extensive number of numerical simulations the transition 
from $2D$ HD turbulence to $2D$ MHD turbulence, varying systematically the magnetic forcing and the involved Reynolds numbers
that allowed us to identify different phases of the turbulent state of 2D MHD. 
Our findings are summarized in figure \ref{fig:Conclusion}.
The top panel shows the variation of $\epsilon_{_E}^-/\epsilon_{_E}$ and $\epsilon_{_A}^-/\epsilon_{_A}$ as a function of
the control parameter $\mu_f$ for fixed $Re^\pm\gg 1, k_f\,L\gg1$. 
For small values of $\mu_f$ the first phase of 2D MHD is met where
the magnetic field is advected passively without altering the hydrodynamics. 
Thus energy cascades inversely and the square vector potential cascades forward.
The magnetic field becomes active first in the small scales when $\mu_f$ is larger than  $\mu_{_{NL}} \propto \left( Re^+ \right)^{-1/n}$.
In this phase, part of the energy cascades forward to the small scales and part of the energy still cascades inversely to the large scales,
while the cascade of the square vector potential remains forward.
Further increasing $\mu_f$ we reach the critical point $\mu_{_Ec}\propto \left( Re^+ \right)^{-1/{2n}}$ where the inverse energy cascade stops and all energy cascades forward.
For slightly larger values we meet the second critical point $\mu_{_Ac}$ that also scales as $ \left( Re^+ \right)^{-1/{2n}} $ and marks the beginning of the inverse cascade of $A$.
Between these two points  $\mu_{_Ec} < \mu_f < \mu_{_Ac}$ both fluxes are strictly forward.
For values of $\mu_f$ larger than $\mu_{_Ac}$ the phase of 2D MHD turbulence exists where the energy cascade is forward and
part of the square vector potential cascades inversely and part forward. 
As $\mu_f$ tends to very large values we expect that forward cascade of $A$ will tend asymptotically to zero 
and $A$ cascades strictly inversely. 
%
%
The different phases are shown in the $\mu_f$ and $Re^+$ parameter space in the lower panel of figure \ref{fig:Conclusion}.
In the phase diagram $Re^{-}$ and $k_fL$ are assumed to be very large or infinity so that the transition clearly exhibits critical behavior 
with well defined critical values of $\mu_f$. 
The three different phases of turbulence are marked on this diagram and are separated by power laws of the form $\mu_f\propto (Re^+)^\gamma$.
Whether $\mu_{_Ec}$ and $\mu_{_Ac}$ converge to a single point as $Re^+$ tends to infinity remains to be seen in future investigations.
 
Our work also indicates that close to the two critical points the dependence of the injection rates and the fluctuation amplitudes
do not necessarily have a smooth behavior allowing the possibility that their derivatives with respect to $\mu_f$ to diverge or be 
noncontinuous. This is not unusual of course in phase transitions where different moments of the fluctuations, or susceptibilities 
scale as non-integer power-laws with the deviation from criticality \cite{stanley1987introduction}. To test such a prospect
and measure these exponents precisely, many runs close to the critical points with both high $Re^+$ and $Re^-$ are needed. 
Unfortunately  current computational limitations restrict us from performing a study in such detail. 

Finally our study indicated that criticality arose from the balance of two counter-cascading processes one driven by the
hydrodynamics (inverse) and one driven by the magnetic field (forward). Such counter-cascading mechanisms also  
exist in other systems exhibiting variable inverse cascade such as rotating and stratified flows \cite{deusebio2014dimensional, sozza2015dimensional}
It remains to be seen if these mechanisms also lead to critical behavior.


\begin{acknowledgments}
This work was granted access to the HPC resources of GENCI-CINES (Project
No.x2014056421, x2015056421)
and MesoPSL financed by the Region Ile de France and the project Equip\@Meso
(reference ANR-10-EQPX-29-01). KS acknowledges support from the LabEx ENS-ICFP:
ANR-10-LABX-0010/ANR-10-IDEX-0001-02 PSL and from {\'E}cole Doctorale {\^I}le de France (EDPIF). 
\end{acknowledgments}

\appendix


\bibliography{apssamp}

\end{document}